\documentclass[envcountsame,runningheads]{llncs}
\usepackage{times}
\usepackage{mathptm}
\usepackage{eucal}
\usepackage{a4}
\usepackage{cite}
\usepackage{amsmath}
\usepackage{amssymb}
\usepackage{amscd}
\usepackage{xspace}
\DeclareMathOperator*{\Freeprod}{\raisebox{-1.4ex}[1ex][0ex]{\text{\huge*}}}
\newcommand{\freeprod}{*}
\newcommand{\range}{\operatorname{range}}

\newcommand{\SL}{\operatorname{SL}}
\newcommand{\IR}{\mathbb{R}}
\newcommand{\R}{\mathbb{R}}
\newcommand{\IC}{\mathbb{C}}
\newcommand{\IA}{\mathbb{A}}
\newcommand{\IB}{\mathbb{B}}
\newcommand{\IQ}{\mathbb{Q}}
\newcommand{\IF}{\mathbb{F}}
\newcommand{\IZ}{\mathbb{Z}}
\newcommand{\IH}{\mathbb{H}}
\newcommand{\IK}{\mathbb{K}}
\newcommand{\IN}{\mathbb{N}}
\newcommand{\N}{\mathbb{N}}
\newcommand{\IM}{\mathbb{M}}
\newcommand{\IP}{\mathbb{P}}
\newcommand{\IS}{\mathbb{S}}

\newcommand{\IO}{\mathbb{O}}
\newcommand{\IX}{\mathbb{X}}

\newcommand{\calH}{\mathcal{H}}

\newcommand{\NP}{\mathcal{NP}}
\newcommand{\person}[1]{\textsc{#1}}
\newcommand{\aname}[1]{\textsf{#1}}
\newcommand{\BSS}{{BSS}\xspace}
\newcommand{\BCSS}{{BSS}\xspace}
\newcommand{\SAG}{algebraically generated\xspace}
\newcommand{\SAE}{algebraically enumerated\xspace}
\newcommand{\SAP}{algebraically presented\xspace}
\newcommand{\SAEP}{algebraically enumerated\slash presented\xspace}
\newcommand{\SAGEP}{algebraically generated\slash enumerated\slash presented\xspace}
\newcommand{\mycite}[2]{\cite[\textsc{#1}]{#2}}
\newcommand{\COMMENTED}[1]{}

\spnewtheorem{observation}[theorem]{Observation}{\bfseries}{\itshape}
\spnewtheorem{fact}[theorem]{Fact}{\bfseries}{\itshape}
\newtheorem{deff}[theorem]{Definition\protect\footnotemark}
\newtheorem{scholiumf}[theorem]{Scholium\footnotemark}
\spnewtheorem{myclaim}[theorem]{Claim}{\bfseries}{\itshape}
\spnewtheorem{myquestion}{Question}{\bfseries}{\itshape}
\newcommand{\ri}{\IR^{\infty}}
\newcommand{\nor}{\text{\rm n}} %generated normal subgroup
\IfFileExists{\jobname.aux}{}{\newlabel{myFn}{{0}{0}}\newlabel{f:Constants}{{0}{0}}\newlabel{f:Intersect}{{0}{0}}}%

\begin{document}
\title{Real Computational Universality: \\
The Word Problem for a class of groups \\ with infinite presentation}
\titlerunning{Computational Universality of the Word Problem for Real Groups}
\author{Klaus Meer\inst{1}\thanks{partially supported 
by  the IST Programme of the European
Community, under the PASCAL Network of Excellence, IST-2002-506778
and by the Danish Natural Science Research Council SNF.
This publication only reflects the authors' views.
Part of the work has been done during a sabbatical which K. Meer
spent at the Forschungsinstitut f\"ur Diskrete Mathematik 
at the university of Bonn, Germany.
The hospitality during that stay is gratefully acknowledged.}
\and 
Martin Ziegler\inst{2}\thanks{supported by \textsf{DFG} (project \texttt{Zi1009/1-1})
and by \textsf{JSPS} (ID \texttt{PE\,05501}).}}%
\authorrunning{K.~Meer, M.~Ziegler}
\institute{IMADA,
Syddansk Universitet,
Campusvej 55,\\ 5230 Odense M, Denmark
(\email{meer@imada.sdu.dk})
\and
Japan Advanced Institute Of Science and Technology,
\\ and University of Paderborn (\email{ziegler@upb.de})
} 
\maketitle
\begin{abstract}
The word problem for discrete groups is well-known to be undecidable
by a Turing Machine; more precisely, it is reducible both to and from
and thus equivalent to the discrete Halting Problem.

The present work introduces and studies
a real extension of the word problem
for a certain class of groups
which are presented as quotient groups of a free group and a
normal subgroup. Most important, these groups may be 
generated by \emph{un}countably many generators
with index running over certain sets of real numbers.
This includes many mathematically important groups which 
are not captured by the finite framework of the classical word problem.

Our contribution extends computational group theory from the 
discrete to the Blum-Shub-Smale (\BSS) model of real number computation.
We believe this to be an interesting step towards
applying \BSS theory, in addition to semi-algebraic geometry, 
also to further areas of mathematics.

The main result establishes the word problem for such groups to be 
not only semi-decidable (and thus reducible \emph{from}) but also 
reducible \emph{to} the Halting Problem for such machines. 
It thus provides the first non-trivial example of a
problem \emph{complete}, that is, computationally universal 
for this model.
\end{abstract}
%%%%%%%%%%%%%%%%%%%%%%%%%%%%%%%%%%%%%%%%%%%%%%%%%%%%%%%%%%%%%%%%%%%%%%
\section{Introduction}
In 1936, \person{Alan M. Turing} introduced the now so-called
Turing Machine and proved the associated Halting Problem $H$,
that is the question of termination of a given such machine $M$,
to be undecidable. On the other hand
simulating a machine $M$ on a Universal Turing Machine
establishes $H$ to be semi-decidable.
In the sequel, several other problems $P$ were also revealed
semi-, yet un-decidable. Two of them,
\aname{Hilbert's Tenth} and the \aname{Word Problem} for groups,
became particularly famous, not least because they arise and
are stated in purely mathematical terms whose relation to
computer science turned out considerable a surprise.
The according undecidability proofs both proceed by constructing
from a given Turing Machine $M$ an instance $x_{M}$ of the problem
$P$ under consideration such that $x_{M}\in P$ iff $M$ terminates;
in other words, a reduction from $H$ to $P$. As $P$ is
easily seen to be semi-decidable this establishes,
conversely, reducibility to $H$ and thus Turing-completeness
of $P$.
%In fact it had been an open question from 1944 \cite{Post} to 1957
%whether \emph{every} semi- yet undecidable problem $P$ is Turing-complete
%in the sense of being reducible from $H$. It was resolved to the
%negative independently by \person{Friedberg} and \person{Muchnik}.

Turing Machines are still nowadays, 70 years after their
introduction, considered the appropriate model of computation for
discrete problems, that is, over bits and integers.
For real number problems of Scientific Computation as for
example in Numerics, Computer Algebra, and Computational Geometry
on the other hand,
several independent previous formalizations were in 1989 subsumed
in a real counterpart to the classical Turing Machines
called the Blum-Shub-Smale, for short \BSS model \cite{BSS,BCSS}. 
It bears many structural similarities
to the discrete setting like for example the existence
of a Universal Machine
%, the notion of ${\cal NP}$--completeness,
or the undecidability of the
associated real Halting Problem $\IH$, that is the
question of termination of a given \BSS-machine $\IM$.

Concerning \BSS-complete problems $\IP$ however, not many are known
so far. The Turing-complete ones for example and, more generally,
any discrete problem becomes decidable over the reals
\mycite{Example~\S1.6}{BSS}; and \emph{extending} an undecidable
discrete problem to the reals generally does not work either:

\begin{example} \label{x:Hilbert}
\aname{Hilbert's Tenth Problem (over $R$)} is the task of deciding,
given a multivariate polynomial equation over $R$, 
whether it has a solution in $R$. 
For integers $R=\IZ$, this problem has been
proven (Turing-)undecidable \cite{Matiyasevich}.
For reals $R=\IR$ however, it \emph{is}
(\BCSS-)decidable by virtue of 
\person{Tarski}'s \aname{Quantifier Elimination} 
\cite[top of p.97]{BCSS}.%
\qed\end{example}
%%%%%%%%%%%%%%%%%%%%%%%%%%%%%%%%%%%%%%%%%
\subsection{Relation to Previous Works}%
Provably undecidable problems over the reals, such as
the \aname{Mandelbrot Set} or the rationals $\IQ$ are supposedly
(concerning the first) or, concerning the latter, have actually been established
\cite{REALPOST}
\emph{not} reducible from and thus strictly easier than $\IH$.
In fact the only \BSS-complete $\IP$ essentially differing from $\IH$
we are aware of is a certain countable existential theory in the
language of ordered fields \mycite{Theorem~2.13}{Cucker}.

The present work closes this structural gap by presenting
a real generalization of the word problem for groups
and proving it to be reducible both from and to the
real Halting Problem. 
On the way to that, we significantly extend notions from 
classical and computational (discrete, i.e.) combinatorial 
group theory to the continuous setting of \BSS-computability.
Several examples reveal these new notions as mathematically
natural and rich. 
They bear some resemblance to certain
recent presentations of continuous fundamental
groups from topology \cite{Hawaii} where, too,
the set of generators (`alphabet') is allowed to be infinite
and in fact of continuum cardinality. There
however words generally have transfinite length
whereas we require them to consist of only
finitely many symbols.

We find our synthesis of 
computational group theory and real number computability to
also differ significantly
from the usual problems studied in the \BSS model which
typically stem from semi-algebraic geometry. 
Indeed, the papers dealing 
with groups $G$ in the \BSS setting \cite{Bourgade,Gassner,Prunescu}
treat such $G$ as underlying structure of the computational model,
that is, not over the reals $\IR$ and its arithmetic.
A rare exception, 
\person{Derksen}, \person{Jeandel}, and \person{Koiran}
do consider \BCSS-decidability (and
complexity) of properties of a real group \cite{Emmanuel};
however they lack completeness results.
Also, their group is not fixed nor presented
but given by some matrix generators.
For instance, finiteness of the multiplicative subgroup of $\IC$
generated by $\exp(2\pi i/x)$, $x\in\IR$, is equivalent to $x\in\IQ$
and thus undecidable yet not reducible from $\IH$ \cite{REALPOST};
whereas any fixed such group is isomorphic either to $(\IZ,+)$ 
or to $(\IZ_n,+)$ for
some $n\in\IN$ and has decidable word problem
(Examples~\ref{x:Circle} and \ref{x:Torus}).

%%%%%%%%%%%%%%%%%%%%%%%%%%%%%%%%%%%%%%%%
\subsection{Overview}
Our work is structured as follows. 
In Section~\ref{secBSS} we recall basic notions
of real number computation. Section~\ref{s:Groups} starts with a review of the classical
word problem in \emph{finitely} presented groups. Then we introduce
real counterparts called \SAP groups, the core objects of our interest.
We give some guiding examples of mathematical groups that fit into this
framework. 
The word problem for these groups is defined and shown
to be semi-decidable in the \BSS model of computation over 
the reals. Section~\ref{s:Hardness} proves our main result:
We recall basic concepts from algebra used in the analysis of the word problem
(Section~\ref{s:Basics})
like Higman-Neumann-Neumann (for short: HNN) extensions and Britton's Lemma
(Section~\ref{s:Effectivity}).
It follows the concept of a benign subgroup (Section~\ref{s:Benign}); 
in the discrete case, 
this notion due to \cite{Higman} relies implicitly on finiteness 
presumptions and thus requires particular care when
generalizing to the continuous case.
Sections~\ref{s:Single} and \ref{s:Final} prove
the paper's central claim:
The real Halting Problem can 
be reduced to the word problem of \SAP real groups.
We close in Section~\ref{s:Conclusion} with some conclusions.

The paper tries to be self-contained for complexity theorists.
This especially holds with respect to the presentation of some concepts
from combinatorial group theory. It is certainly recommended to study
the related material from original sources. 
In particular, we found the books by \person{Rotman} \cite{Rotman} and 
by \person{Lyndon} and \person{Schupp} \cite{Lyndon} extremely helpful.

%%%%%%%%%%%%%%%%%%%%%%%%%%%%%%%%%%%%%%%%%%%%%%%%%%%%%%%%%%%%%%%%%%%%%%%%
\section{\BSS-Machines and the Real Halting Problem} \label{secBSS}

This section summarizes very briefly the main ideas
of real number computability theory. For a more detailed
presentation see \cite{BCSS}.

Essentially a (real) \BSS-machine can be considered as a \textsf{Random Access Machine}
over $\IR$ which is able to perform the basic arithmetic operations  at
unit cost and which registers can hold arbitrary real numbers.
Its inputs are thus finite sequences over $\IR$ of possibly
unbounded length.

\begin{definition}{\rm\cite{BSS}}   \label{defBCSS}
\begin{enumerate}
\item[a)]
Let $\IX \subseteq \ri := \biguplus_{d \in \IN} \IR^d $, i.e. a set of
finite sequences of real numbers.
Its \textsc{dimension}, $\dim(\IX)$, is the smallest
$D\in\IN$ such that $\IX\subseteq\bigoplus_{d\leq D}\IR^d$;
$\dim(\IX)=\infty$ if no such $D$ exists.
\item[b)]
A {\sc \BSS-machine $\IM$
over $\IR$ with admissible input set $\IX$} is given by a finite set
$I$ of
instructions labeled by $ 1,\ldots,N . $ A configuration of $\IM$
is a quadruple $(n,i,j,\bar y) \in I \times \IN \times \IN \times \ri.$ Here, 
$n$ denotes the currently executed instruction, $i$ and $j$ are used as
addresses (copy-registers) and $\bar y$ is the actual content of the registers
of $\IM$. The initial configuration of $\IM$'s computation on input $\bar x \in \IX$
is $ (1,1,1,\bar x) $ . If $ n = N $ and the actual configuration is
$(N,i,j,\bar y) $, the computation stops with output $\bar y$. 
The instructions $\IM$ is allowed to perform are of the following types:
\smallskip
\begin{description}
\item[computation:\ ] $n: y_s\leftarrow y_k \circ_n y_l$,
where $\circ_n \in \{+,-,\times,\div\}$;\quad or \newline
~$n: y_s \leftarrow \alpha$ for some $ \alpha \in \IR \ . $\newline
The register $\#s$ will get the value $ y_k \circ_n y_l $ or $ \alpha$, 
respectively. 
All other register-entries remain unchanged.
The next instruction will be $n+1$;
moreover, 
the copy-register $i$ is either incremented by one, replaced by $0$, or
remains unchanged. The same holds for copy-register $j$.
\item[branch:\ ] $n$: {\bf if $y_0\geq 0$ goto $\beta(n)$ else goto $n+1$.}
According to the answer of the test the next instruction is determined
(where $ \beta(n) \in I ).$ All
other registers are not changed.
\item[copy:\ ] $n: y_i\leftarrow y_j$, i.e. the content of
the ``read"-register is copied into the ``write"-register. The
next instruction is $n+1$; all other registers remain unchanged.
\end{description}
\item[c)] The size of an $\bar x \in \IR^d$ 
%$x=(x_1,\ldots,\underbrace{x_k}_{\neq 0},0,\ldots) 
%\in \ri $ 
is $size_{\IR}(\bar x)=d$. The cost of any of the
above operations is $1$. The cost of a computation
is the number of operations performed until the machine halts.

\item[d)] For some $\IX \subseteq \ri$ we call a function $f :\IX \to \ri$ 
\mbox{(\BSS-)}computable iff it is realized by a \BSS machine over admissible
input set $\IX$. Similarly, a set $\IX \subseteq \ri$ is 
decidable in $\ri$ iff its characteristic function is computable. 
$\IX$ is called a decision problem or a language over $\ri.$%
\item[e)]
A \BSS oracle machine using an oracle set $\IO \subseteq \ri$
is a \BSS machine with an additional type of node called oracle node.
Entering  such a node the machine can ask the oracle whether a previously
computed element $\bar y\in\ri$ belongs to $\IO.$ The oracle gives the correct
answer at unit cost.
\end{enumerate}
\end{definition}
A real  Halting Problem now can be 
defined straightforwardly as well.

\begin{definition}\label{Definition:Halting-Problem}
The \emph{real Halting Problem} $\IH$ 
is the following decision problem. Gi\-ven 
the code $c_{\IM} \in \ri$ of a \BSS machine $\IM$, 
%together with an $x \in \ri$, 
does $M$ ter\-mi\-nate its computation (on input $0$) ?
\end{definition}
Both the existence of such a coding for \BSS machines and 
the undecidability of $\IH$ in the \BSS model were shown in \cite{BSS}.

%%%%%%%%%%%%%%%%%%%%%%%%%%%%%%%%%%%%%%%%%%%%%%%%%%%%%%%%%%%%%%%%%%%%%%%%
\section{Word-Problem for Groups} \label{s:Groups}
Groups occur ubiquitously in mathematics,
and having calculations with and in them 
handled by computers constitutes an important
tool both in their theoretical investigation
and in practical applications as revealed
by the flourishing field of \emph{Computational
Group Theory} \cite{DIMACSa,DIMACSb,Handbook}.
Unfortunately already the simplest question,
namely equality ~`$a=b$'~ of two elements $a,b\in G$
is in general undecidable for groups $G$ reasonably
presentable to a digital computer, that is, 
in a finite way --- the celebrated result obtained
in the 1950ies
independently by 
\person{Novikov} \cite{Novikov} and
\person{Boone} \cite{Boone}.
In the canonical model of real number decidability\footnote{%
We remark that 
in the other major and complementary model of
real number computation, decidability makes no sense
as it corresponds to evaluating a characteristic
and thus discontinuous function which is uncomputable
due to the so-called Main Theorem of Recursive
Analysis \mycite{Theorem~4.3.1}{Weihrauch}.}
on the other hand, \emph{every} discrete problem
$L\subseteq\Sigma^*$ is solvable
\mycite{Example~\S1.6}{BSS}, rendering the
word problem for finitely presented groups
trivial.

However, whenever we deal with computational questions involving
groups of real or complex numbers, the Turing model seems
not appropriate anyway. As an example take the 
unit circle in $\IR^2$ equipped with complex multiplication.
There is a clear mathematical intuition how to compute
in this group; such computations can be formalized in the \BSS model.
We thus aim at a continuous counterpart to the discrete class
of finitely presented groups for which the word problem
is universal for the \BSS model.

\medskip
After recalling basic notions related to the (classical) word problem of finitely
presented groups (Section~\ref{s:DiscreteGroups})
we introduce in Section~\ref{s:RealGroups}
the larger class of \SAP real groups.
Section~\ref{s:Examples} gives several examples 
showing how this new class covers natural groups 
occurring in mathematics. 
Next (Section~\ref{s:Semidecidable}) we establish 
semi-decidability of the word problem for \SAP groups,
that is, reducibility to the Halting Problem $\IH$
in the real number model of Blum, Shub, and Smale.
Our main result then proves the existence of 
\SAP groups
for which the word problem is reducible from $\IH$;
this covers the entire Section~\ref{s:Hardness}.

%%%%%%%%%%%%%%%%%%%%%%%%%%%%%%%%%%%%%%%%%%
\subsection{The Classical Setting}
\label{s:DiscreteGroups}

Here, the setting for the classical word problem is briefly recalled.
A review of the main algebraic concepts needed in our proofs
is postponed to Section~\ref{s:Hardness}.

\begin{definition}\label{d:Groups0}
\begin{enumerate}
\item[a)]
Let $X$ be a set. The \emph{free group generated by $X$},
denoted by $F=(\langle X\rangle,\circ)$ or more briefly
$\langle X\rangle$, is the set
$(X\cup X^{-1})^*$ of all finite sequences
$\bar w=x_1^{\epsilon_1}\cdots x_n^{\epsilon_n}$ with $n\in\IN$,
$x_i\in X$, $\epsilon_i\in\{-1,+1\}$, equipped with
concatenation $\circ$ as group operation subject
to the rules 
\begin{equation} \label{e:Rule0}
  x\circ x^{-1}\quad=\quad1\quad=\quad x^{-1}\circ x \qquad \forall x\in X
\end{equation}
where $x^1:=x$ and where $1$ denotes the empty word,
that is, the unit element.
\item[b)]
For a group $H$ and $W\subseteq H$, denote by
\[ \langle W\rangle_{H} \;:=\;\big\{w_1^{\epsilon_1}\cdots w_n^{\epsilon_n}:
n\in\IN, w_i\in W, \epsilon_i=\pm1\big\} \]
the subgroup of $H$ generated by $W$.
The \emph{normal} subgroup of $H$ generated by $W$ is%
\[ \langle W\rangle_{H\!\nor} \;:=\; 
%\big\{ h_1^{}\cdot \bar w_1\cdot h_1^{-1}\,
%\cdot\,h_2^{}\cdot\bar w_2\cdot h_2^{-1}\,\cdots\,h_n^{}\cdot\bar w_n\cdot h_n^{-1}:
%  n\in\IN, h_i\in H, \bar w_i\in \langle W\rangle_H\big\}
\langle \{ h\cdot w\cdot h^{-1} : h\in H, w\in W\}\rangle_{H} 
\enspace . \]
For $h\in H$, we write $h/W$ for its $W$--coset 
$\{h\cdot w:w\in\langle W\rangle_{H\!\nor}\}$
of all $g\in H$ with $g\equiv_W h$.%
\item[c)]
Fix sets $X$ and $R\subseteq\langle X\rangle$ and
consider the quotient group $G:=\langle X\rangle/\langle R\rangle_{\nor}$,
denoted by $\langle X|R\rangle$, of all $R$--cosets of $\langle X\rangle$.

If both $X$ and $R$ are finite, the tuple $(X,R)$
will be called a \emph{finite presentation} of $G$;
if $X$ is finite and $R$ recursively enumerable (by a Turing machine,
that is in the discrete sense; equivalently: semi-decidable), 
it is a \emph{recursive\footnote{This notion seems misleading
as $R$ is in general \emph{not} recursive; nevertheless
it has become established in literature.} presentation};
if $X$ is finite and $R$ arbitrary, 
$G$ is \emph{finitely generated}.
\end{enumerate}
\end{definition}
Intuitively, $R$ induces further rules ``$\bar w=1$, $\bar w\in R$''
in addition to Equation~(\ref{e:Rule0});
put differently, distinct words $\bar u,\bar v\in\langle X\rangle$
might satisfy $\bar u=\bar v$ \emph{in $G$}, that is, by virtue of $R$.
Observe that the rule 
  ``$\displaystyle w_1^{\epsilon_1}\cdots w_n^{\epsilon_n}=1$'' 
induced by an element 
$\displaystyle\bar w=(w_1^{\epsilon_1}\cdots w_n^{\epsilon_n})\in R$ 
can also be applied as
``$\displaystyle
  w_1^{\epsilon_1}\cdots w_k^{\epsilon_k}=w_n^{-\epsilon_n}\cdots w_{k+1}^{-\epsilon_{k+1}}$''.
\addtocounter{theorem}{-1}
\begin{definition}[continued]
\begin{enumerate}
\item[d)]
The \emph{word problem for $\langle X|R\rangle$} is
the task of deciding, given $\bar w\in\langle X\rangle$,
whether $\bar w=1$ holds in $\langle X|R\rangle$.
\end{enumerate}
\end{definition}
The famous work
of Novikov and, independently, Boone
establishes the word problem for finitely presented
groups to be Turing-complete:
\begin{fact} \label{f:Groups0}
\begin{enumerate}
\item[a)]
For any finitely presented group $\langle X|R\rangle$,
its associated word problem is semi-de\-ci\-da\-ble
(by a Turing machine).
\item[b)]
There exists a finitely presented
group $\langle X|R\rangle$ whose associated word problem
is many-one reducible by a Turing machine
from the discrete Halting Problem $H$.
\qed\end{enumerate}
\end{fact}
Of course, a) is immediate.
For the highly nontrivial Claim~b), 
see e.g. one of \cite{Boone,Novikov,Lyndon,Rotman}.
\begin{example} \label{x:InfiniteGrp}
\quad $\displaystyle
\calH \;\;:=\;\;
\big\langle \{a,b,c,d\}\,\big|\,\{a^{-i}ba^i=c^{-i}dc^i: i\in H\}\big\rangle$
\\ is a \emph{recursively} presented group with word problem
reducible from $H$; compare the proof of
\mycite{Theorem~\textsection IV.7.2}{Lyndon}.
\qed\end{example}
In order to establish Fact~\ref{f:Groups0}b), we
need a \emph{finitely} presented group.
This step is provided by the remarkable
\begin{fact}[Higman Embedding Theorem] \label{f:Higman}
Every recursively presented group
can be embedded in a finitely generated one.%
\end{fact}
\begin{proof}
See, e.g., \mycite{Section~\textsection IV.7}{Lyndon} or
\mycite{Theorem~12.18}{Rotman}.
\qed\end{proof}
Fact~\ref{f:Higman} asserts the word problem %of $\calH$
from Example~\ref{x:InfiniteGrp}
to be in turn reducible to that of the finitely 
presented group $\calH$ is embedded into, because
any such embedding is automatically effective:%
\begin{observation} \label{o:Embedding}
Let $G=\langle X\rangle/\langle R\rangle_{\nor}$
and $H=\langle Y\rangle/\langle S\rangle_{\nor}$
denote finitely generated groups
and $\psi:G\to H$ a homomorphism.
Then, $\psi$ is (Turing-) computable in the sense that
there exists a computable homomorphism
$\psi':\langle X\rangle\to\langle Y\rangle$
such that 
$\psi(\bar x)\in\langle S\rangle_{\nor}$
whenever $\bar x\in\langle R\rangle_{\nor}$;
that is, $\psi'$ maps $R$-cosets to $S$-cosets 
and makes the following diagram commute:
\begin{equation} \label{e:Embedding}
\begin{CD}
\langle X\rangle @>>\psi'> \langle Y\rangle \\
@VVV @VVV\\
\langle X\rangle/\langle R\rangle_{\nor}
@>\psi>> \langle Y\rangle/\langle S\rangle_{\nor}
\end{CD}
\end{equation}
\end{observation}
Indeed, due the homomorphism property, $\psi$ is uniquely 
determined by its values on the finitely many generators 
$x_i\in X$ of $G$,
that is, by $\psi(x_i)=\bar w_i/\langle S\rangle_{\nor}$
where $\bar w_i\in\langle Y\rangle$. Setting (and storing
in a Turing Machine) 
$\psi'(x_i):=\bar w_i$ yields the claim.

%%%%%%%%%%%%%%%%%%%%%%%%%%%%%%%%%%%%%%%%%%
\subsection{Presenting Real Groups} \label{s:RealGroups}
Regarding that the \BSS-machine is the natural extension
of the Turing machine from the discrete to the reals,
the following is equally natural a generalization
of Definition~\ref{d:Groups0}c+d):
\begin{deff} \label{d:Groups1}
\footnotetext{Making a definition bears similarities to procreation:
It may become a lasting source of happiness and joy;
but if performed imprudently, you might later regret it deeply.}
Let $X\subseteq\ri$ and 
$R\subseteq\langle X\rangle\subseteq\!\!\!\footnotemark\,\ri$.%
\footnotetext{Most formally, $R$ is 
a set of \emph{vectors of vectors} of varying lengths. 
However by suitably encoding delimiters,
we shall regard $R$ as effectively embedded into
\emph{single} vectors of varying lengths.}
The tuple $(X,R)$ is called a \emph{presentation}
of the \emph{real group} $G=\langle X|R\rangle$.
This presentation is \emph{\SAG} if $X$ 
is \BSS-decidable and $X\subseteq\IR^N$ for some $N\in\IN$.
$G$ is termed \emph{\SAE}
if $R$ is in addition \BSS semi-decidable;
if $R$ is even \BSS-decidable,
call $G$ \emph{\SAP.}
%For $X\subseteq\IR^d\times\IZ^k$,
%$d$ denotes the \emph{dimension} of $G$.
%
The \emph{word problem} for the presented real group 
$G=\langle X|R\rangle$ is
the task of \BSS-deciding, given $\bar w\in\langle X\rangle$,
whether $\bar w=1$ holds in $G$.
\end{deff}
The next table summarizes the correspondence between the
classical discrete and our new real notions.
\begin{center}
\begin{tabular}{l@{\;\;}|@{\;\;}l}
\hspace*{\fill} Turing\hspace*{\fill}  & \hspace*{\fill} \BCSS\hspace*{\fill} 
\\ \hline
finitely generated  & \SAG \\
recursively presented & \SAE \\
finitely presented & \SAP
\end{tabular}
\end{center}
%A few comments are in place.
\begin{remark} \label{r:Real}
\begin{enumerate}
\item[a)]
Although $X$ inherits from $\IR$ algebraic structure
such as addition $+$ and multiplication $\times$, 
the Definition~\ref{d:Groups0}a) of the free group
$G=(\langle X\rangle,\circ)$ considers $X$ as a plain set only.
In particular, (group-) inversion in $G$ must not be confused with
(multiplicative) inversion:
$5\circ\tfrac{1}{5}\not=1=5\circ 5^{-1}$ for $X=\IR$.
This difference may be stressed notationally by writing 
`abstract' generators $x_{\bar a}$ indexed
with real vectors $\bar a$; here, `obviously'
$x_5^{-1}\not=x_{\scriptscriptstyle1/5}^{}$.
%$x_{\bar a}=x_{\bar b}:\Leftrightarrow\bar a=\bar b$.
\COMMENTED{
While Definition~\ref{d:Groups0}c) requires the set $X$ of
generators to be finite, it must in the real setting be
a finite-\emph{dimensional} subset of $\ri$. 
Asserting this condition is the vital part of our 
construction (cmp. e.g. Observation~\ref{o:Dimension})
whereas without it, \BSS-hardness of the real word
problem becomes trivial (see Example~\ref{x:InfiniteDim}).

Definition~\ref{d:Groups1} neglects possible discrete 
contributions to the generators; this takes into account
that integers, as opposed to reals, admit an effective pairing 
function $\langle\,\cdot\,,\,\cdot\,\rangle:\IN\times\IN\to\IN$
which in turn yields a bicomputable bijection
$\IR\times\IN\to\IR$ like
$(x,n)\mapsto\langle\lfloor x\rfloor,n\rangle+(x-\lfloor x\rfloor)$.
}
\item[b)]
Isomorphic (that is, essentially
identical) groups 
$\langle X|R\rangle\cong\langle X'|R'\rangle$
may have different presentations $(X,R)$ and $(X',R')$;
see Section~\ref{s:Examples}. Even when $R=R'$,
$X$ need not be unique!
Nevertheless we adopt from literature 
such as \cite{Lyndon} the convention\footnote{\label{myFn}%
This can be justified with respect to the solvability
of the word problem in the case of 
\emph{finite} presentations and \emph{non}uniformly
by virtue of \person{Tietze}'s Theorem
\cite[\textsc{Propositions~\textsection II.2.1}
and \textsc{\textsection II.2.2}]{Lyndon}.} of speaking of
``the group $\langle X|R\rangle$'',
meaning a group with presentation $(X,R)$.

This however requires some care, for instance when
$\bar w$ is considered (as in Definition~\ref{d:Groups0}d) 
both an element of 
$\langle X\rangle$ and of $\langle X|R\rangle$!
For that reason we prefer to write
$\langle W\rangle_H$ rather than, e.g.,
$\operatorname{Gp}(W)$: to indicate in which 
group we consider a subgroup to be generated.
\item[c)]
By means of Turing-computable pairing functions
$\langle\,\cdot\,,\,\cdot\,\rangle:\IN\times\IN\to\IN$
like $(k,\ell)\mapsto 2^k\cdot(2\ell+1)$,
vectors of integers are effectively en- and decodable
into one single component. Although a \emph{real} pairing
function $\IR\times\IR\to\IR$ can\emph{not} be
\BSS-computable, the set $X$ of generators of an
\SAG group may easily be achieved to live in 
$\IR^{\pmb{1}}\times\IN$ by effectively proceeding
from a vector generator $x_{(r_1,\ldots,r_d)}$ to a
word $x_{r_1}\circ y_1\circ x_{r_2}\circ y_2\cdots x_{r_d}\circ y_d$
over $\IR\times\IN\cong\{x_r:r\in\IR\}\cup\{y_i:i\in\IN\}$
and adjusting $R$ accordingly.
\end{enumerate}
\end{remark}
For a \BSS-machine to read or write a word 
$\bar w\in\langle X\rangle=(X\cup X^{-1})^*$ of course means
to input or output a vector
$(w_1,\epsilon_1,\ldots,w_n,\epsilon_n)\in(\IR^N\times\IN)^n$.
In this sense, the Rules~(\ref{e:Rule0}) 
implicit in the free group are obviously decidable
and may w.l.o.g. be included in $R$.%
%%%%%%%%%%%%%%%%%%%%%%%%%%%%%%%%%%%%%%%%%%%%%%%%%%
\subsection{Examples} \label{s:Examples}
\begin{example} \label{x:Discrete}
Every finite \emph{or recursive} presentation 
is an algebraic presentation.
Its word problem is \BSS-decidable.
\end{example}
As long as $X$ in Definition~\ref{d:Groups0}c) is at most countable,
so will be any group $\langle X|R\rangle$. Only proceeding
to real groups as in Definition~\ref{d:Groups1} can
include many interesting uncountable groups in mathematics.
\begin{example} \label{x:Circle}
Let $\IS$ denote the unit circle in $\IC$ with complex multiplication.
The following is an 
algebraic presentation $\langle X|R_1\cup R_2\rangle$ of $\IS$:
\begin{itemize}
\itemsep4pt
\item[\textbullet] $\displaystyle 
  X:=\big\{x_{r,s}:(r,s)\in\IR^2\setminus\{0\}\big\}$, 
\item[\textbullet] $\displaystyle 
  R_1:=\big\{x_{r,s}^{}\circ x_{a,b}^{-1}:(r,s),(a,b)\not=0\wedge rb=sa \wedge ar>0\big\}$,
\item[\textbullet] $\displaystyle 
  R_2:=\big\{x_{r,s}^{}\circ x_{a,b}^{}\circ x_{u,v}^{-1}:(r,s),(a,b),(u,v)\not=0\wedge\\
\hspace*{\fill} \wedge r^2+s^2=1\wedge a^2+b^2=1\wedge u=ra-sb\wedge v=rb+sa\big\}$.
\end{itemize}
Intuitively, $R_1$ yields the identification of (generators whose
indices represent) points lying on the same
half line through the origin. In particular, every $x_{r,s}$
is `equal' (by virtue of $R_1$) to some $x_{a,b}$ of `length'
$a^2+b^2=1$. To these elements, $R_2$ applies and identifies 
$x_{r,s}\circ x_{a,b}$ with $x_{u,v}$ whenever, 
over the complex numbers, it holds $(r+is)\cdot(a+ib)=u+iv$.
\qed\end{example}
Clearly, the presentation of a group need not be unique;
e.g. we also have
$\IS\cong\langle Y|R_2\rangle$ where $Y=\{x_{r,s}:r^2+s^2=1\}$.
Here is a further algebraic presentation of the same group:%
\begin{example} \label{x:Torus}
Let $X:=\{x_t:t\in\IR\}$, $R:=\{x_t=x_{t+1},\, x_t x_s=x_{t+s}:t,s\in\IR\}$.
Then $\langle X|R\rangle$ is a 1D (!) algebraic presentation of
the group $\big([0,1),+\big)$ isomorphic to
$(\IS,\times)$ via $t\mapsto\exp(2\pi i t+ic)$ for any $c\in\IR$.
Yet none of these isomorphisms is \BSS-computable!%
\qed\end{example}
Next consider the group $\SL_2(\IR)$ of real $2\times 2$ matrices
$A$ with $\det(A)=1$. A straight-forward %4-dimensional 
algebraic presentation 
of it is given as $\langle X|R\rangle$ where
$X:=\{x_{(a,b,c,d)}:ad-bc=1\}$ and 
$R:=\{x_{(a,b,c,d)}x_{(q,r,s,t)}=x_{(u,v,w,z)}:
  u=aq+bs\wedge v=ar+bt\wedge w=cq+ds\wedge z=cr+dt\}$.
Here as well as in the above examples, any
group element $\bar w\in\langle X\rangle$
is equivalent (w.r.t. $R$) to an appropriate 
\emph{single} generator $x\in X$. This 
is different %does not hold 
for the following alternative,
far less obvious %and, surprisingly, only \emph{one}-dimensional 
algebraic presentation:%
\begin{example}[Weil Presentation of $\SL_2(\IR)$] \label{x:Weil}
For each $b \in \R$, write
\[
U(b):=\left( \begin{array}{cc} 1 & b \\ 0 & 1 \end{array} \right),
\quad V:= \left( \begin{array}{rr}0&1\\-1&0 \end{array} \right),
\quad 
 S(a):= V \cdot U(\tfrac{1}{a})\cdot V \cdot U(a) \cdot  V \cdot U(\tfrac{1}{a}) 
 \in \SL_2(\IR) \enspace . \]
Let $\displaystyle X=\{x_{U(b)}:b\in\IR\}\cup\{X_V\}$.
Furthermore let $R$ denote the union of the following four 
families of relations (which are easy but tedious to
state formally as subsets of $\langle X\rangle$):
\begin{description}
\itemsep1pt
\item[SL1:] ``\emph{$U(\cdot)$ is an additive homomorphism}'';
\item[SL2:] ``\emph{$S(\cdot)$ is a multiplicative homomorphism}'';
\item[SL3:] ``\emph{$V^2 = S(-1) $}'';
\item[SL4:] ``\emph{$S(a) \cdot U(b) \cdot S(1/a) = U(b a^2) \;\forall a,b$}''.
\end{description}
According to \cite{Lang}, 
$\langle X|R\rangle$
is isomorphic to $SL_2(\R)$ under the natural homomorphism.%
\qed\end{example}
In all the above cases, the word problem
--- in Example~\ref{x:Circle} basically the question whether
$(r,s)=(1,0)$ and in Example~\ref{x:Torus}
whether $t=0$ ---
is decidable. We next illustrate
that, in the real case, different presentations of the same group
may affect solvability of the word problem.
\begin{example} \label{x:Rationals}
The following are presentations $\langle X|R\rangle$ of $(\IQ,+)$:
\begin{enumerate}
\itemsep2pt
\item[a)]
  $\displaystyle X\;=\;\big\{ x_r : r\in\IQ \big\}$,
  \quad $\displaystyle R\;=\;\big\{ x_r x_s=x_{r+s}: r,s\in\IQ\big\}$.
\item[b)]
  $\displaystyle X\;=\;\{ x_{p,q} : p,q\in\IZ, q\not=0\}$,\\
  $\displaystyle R\;=\;\big\{ x_{p,q} x_{a,b}=x_{(pb+aq,qb)}:
    p,q,a,b\in\IZ\big\}\;\cup\;\big\{
   x_{p,q}=x_{(np,nq)}:p,q,n\in\IZ, n\not=0\big\}$.%
\item[c)]
  Let $(b_i)_{i\in I}$ denote an algebraic basis\footnote{%
  That is, as opposed to a Banach space basis, 
  every vector admits a representation as 
  linear combination of \emph{finitely} many out of these
  (here uncountably many) base elements. 
  This is sometimes termed a \emph{Hamel basis},
  a notion we prefer to avoid \mycite{pp.309--310}{Poliakov}.} 
  of the $\IQ$--vector space $\IR$; w.l.o.g. $0\in I$ and $b_0=1$.
  Consider the linear projection $P:\IR\to\IQ$,
  $\sum_i r_i b_i\mapsto r_0$ with $r_i\in\IQ$.
\[ X\;=\; \big\{ x_t:t\in\IR \big\}, \quad
  R\;=\; \big\{ x_t x_s=x_{t+s}:t,s\in\IR\big\}
    \;\cup\;\big\{ x_t=x_{P(t)}:t\in\IR\big\} \enspace . \]
\end{enumerate}
Case~b) yields an algebraic presentation,
a) is not even \SAG but c) is.
The word problem is decidable for a): e.g. by effective
embedding into $(\IR,+)$; and so is it for b) 
although not for c):
$x_t=1\Leftrightarrow P(t)=0$ but both
$P^{-1}(0)=\{\sum_{j\in J} b_j q_j:0\not\in J\subseteq I\text{ finite}, q_j\in\IQ\}$
and its complement are totally disconnected and
uncountable, hence \BSS-undecidable.%
\qed\end{example}
\begin{example} \label{x:Undecidable}
(Undecidable) real membership ``$t\in\IQ$'' 
is reducible to the word problem
of an \SAP real group:
Consider $X=\{x_r:r\in\IR\}$,
$R  = \big\{
  x_{nr}=x_r,%:r\in\IR, %n\in\IN\big\}
%   \;\cup\;\big\{ 
   x_{r+k}=x_k : r\in\IR, n\in\IN,k\in\IZ\big\}$.
Then $x_r=x_0\Leftrightarrow r\in\IQ$;
also, $R\subseteq\IR^2$ is decidable 
because $\IZ\subseteq\IR$ is.%
\qed\end{example}
This however does not establish \BSS-\emph{hardness}
of the real word problem because $\IQ$ is provably
easier than the \BSS Halting Problem $\IH$ 
\cite{REALPOST}. On the other hand, without the 
restriction to \SAP groups (and thus parallel
to Example~\ref{x:InfiniteGrp}), it is easy to find 
a real group with \BSS-hard word problem:

\begin{example} \label{x:InfiniteDim}
Let \ $X\;:=\;\{x_r,y_r:r\in\IR\}\uplus\{s,t\}\cong(\IR\uplus\{\infty\})\times\{1,2\}$.
and $R:=\{\bar v_{\bar r}=\bar w_{\bar r}:\bar r\in\IH\}$ where,
for $\bar r\in\IR^d$, we abbreviated
$\bar v_{\bar r}:=x^{-1}_{r_d}\cdots x^{-1}_{r_1}\cdot s\cdot
x^{}_{r_1}\cdots x^{}_{r_d}$ and
$\bar w_{\bar r}:=y^{-1}_{r_d}\cdots y^{-1}_{r_1}\cdot t\cdot
y^{}_{r_1}\cdots y^{}_{r_d}$. 
In $G:=\langle X|R\rangle$,
it is $\bar v_{\bar r}=\bar w_{\bar r}$ 
iff $\bar r\in\IH$; compare Fact~\ref{f:Nielsen}.
Therefore, $\bar r\mapsto\bar v_{\bar r}\cdot \bar w_{\bar r}$
constitutes a reduction from $\IH$ to the word
problem in $G$. However, $G$ has 
just \emph{semi}-decidable relations.
\qed\end{example}
The construction of an \emph{\SAP} group
with \BSS-complete word problem 
in Section~\ref{s:Hardness}
is the main contribution
of the present work.
%%%%%%%%%%%%%%%%%%%%%%%%%%%%%%%%%%%%%%%%
\subsection{Reducibility \emph{to} the Real Halting Problem} 
\label{s:Semidecidable}
We first show that, parallel to
Fact~\ref{f:Groups0}a), the word problem 
for any \SAE real group is not 
harder than the \BSS Halting Problem. 
\begin{theorem}\label{t:Semidecidable}
Let $G=\langle X|R\rangle$ denote a \SAE real group.
Then the associated word problem is \BSS semi-decidable.
\end{theorem}
Recall that semi-decidability of
$A\subseteq\IK^\infty$
(that is, being a halting set)
is equivalent to recursive enumerability
\[
A=\range(f) \quad\text{ for some computable, partial function }
f:\subseteq\IK^\infty\to\IK^\infty
\]
in the Turing ($\IK=\IF_2$)
as well as the \BSS ($\IK=\IR$) model;
in the latter case by virtue of 
\person{Tarski}'s quantifier elimination \cite{Michaux}.

\begin{lemma} \label{l:DecidableSpan}
For $Y\subseteq\ri$, it holds:
If $Y$ is (semi-)decidable, then so is $\langle Y\rangle$.
\end{lemma}
\begin{proof}
Given a string $\bar w=(y_1,\ldots,y_k)\in\IR^k$,
consider all $2^{k-1}$ partitions of $\bar w$ into non-empty
subwords. For each subword, decide or semi-decide whether
it belongs to $Y\cup Y^{-1}$. Accept iff, for at least one
partition, all its subwords succeed.
\qed\end{proof}

\begin{proof}[Theorem~\ref{t:Semidecidable}]
By Definition~\ref{d:Groups0}b+c), 
$\bar w\equiv 1\Leftrightarrow \bar w\in\langle W\rangle_{\nor}$,
that is, if and only if 
\begin{equation} \label{e:Semidecidable}
\exists n\in\IN\;\;
\exists \bar x_1,\ldots,\bar x_n\in \langle X\rangle\;\;
\exists \bar r_1,\ldots,\bar r_n\in \langle R\rangle: \quad
  \bar w = \bar x_1^{}\bar r_1\bar x_1^{-1} \,\cdot\,
   \bar x_2^{}\bar r_2\bar x_2^{-1} \,\cdots\,
   \bar x_n^{}\bar r_n\bar x_n^{-1} \enspace .
\end{equation}
Since both $X$ and $R$ were required to be semi-decidable,
same holds for $\langle X\rangle$ and $\langle R\rangle$.
This yields semi-decidability of (\ref{e:Semidecidable}).
Indeed, let $f,g:\subseteq\ri\to\ri$ be \BSS-computable with
$\langle X\rangle=\range(f)$ and $\langle R\rangle=\range(g)$;
then it is easy to construct (but tedious to formalize) 
from $f$ and $g$ a \BSS-computable function on $\ri$
ranging over all
$n\in\IN$, all $\bar w\in\langle X\rangle$,
all $\bar x_1,\ldots,\bar x_n\in\langle X\rangle$,
and all $\bar r_1,\ldots,\bar r_n\in\langle R\rangle$.
Compose its output with the decidable test
``$\bar w = \bar x_1^{}\bar r_1\bar x_1^{-1} \cdots
   \bar x_n^{}\bar r_n\bar x_n^{-1}$?'' and,
if successful, return $\bar w$.
This constitutes a function on $\ri$
with range exactly $\langle W\rangle_{\nor}$.
\qed\end{proof}

%%%%%%%%%%%%%%%%%%%%%%%%%%%%%%%%%%%%%%%%%%%%%%%%%%%%%%%%%%%%%%%%%%%%%%%%
\section{Reduction \emph{from} the Real Halting Problem} \label{s:Hardness}
This section proves the main result of the paper
and continuous counterpart to Fact~\ref{f:Groups0}b):
The word problem for \SAP real groups is in general not
only undecidable (cf. Example~\ref{x:Undecidable})
in the \BSS model but in fact as hard as the real Halting Problem.
\begin{theorem} \label{t:Hardness}
There exists an \SAP real group $\calH=\langle X|R\rangle$
such that $\IH$ is \BSS-reducible to the word problem
in $\calH$.
\end{theorem}
We first (Sections~\ref{s:Basics}) review some basics from group 
theory in the context of presented groups;
specifically free products, HNN extensions and \person{Britton}'s Lemma. 
As in the classical reduction from the Turing Halting Problem $H$ to
finitely presented groups in \mycite{Section~\textsection IV.7}{Lyndon} 
(based on ideas of \person{Higman} \cite{Higman} and \person{Valiev} \cite{Valiev}), 
these powerful tools permit a more elegant and abstract treatment
than the elementary approach pursued in, e.g., \mycite{Chapter~12}{Rotman}.
A second major ingredient, \emph{benign} subgroups are recalled 
and generalized to our effective real setting in Section~\ref{s:Benign}.
This requires particular care since many properties heavily exploited
in the discrete case (e.g., that
the homeomorphic image of a finitely generated group is again
finitely generated) are not immediately clear how to carry over
to the reals (Section~\ref{s:Effectivity}). For instance,
a proof for the classical result may exploit \person{Matiyasevich}'s 
famous solution of Hilbert's Tenth Problem, namely a 
Diophantine formulation of $H$ \cite{Matiyasevich}.
This form can be transformed into a straight line program
and further on into a group theoretic one by virtue of 
\person{Higman}'s concept of benign subgroups.
Our general proof strategy is conceptually similar but necessarily
quite different in detail. 
Specifically, lacking a real Diophantine characterization
of $\IH$ (recall Example~\ref{x:Hilbert}),
Section~\ref{s:Single} has to proceed differently,
namely by describing each \emph{fixed} computational path 
of a \BSS machine as a real straight line program, and obtains from that 
a representation as an \emph{effectively} benign real group.
In the final step (Section~\ref{s:Final}), all these groups and their 
embeddings are joined into one single, \SAP one.
%Since this resulting presentation must have finite dimension,
%particular attendance and effort is required for its constituents'
%dimensions to remain bounded. We actually attain dimension 4 (and could
%improve it further down to 2 in case that were considered important).

%%%%%%%%%%%%%%%%%%%%%%%%%%%%%%%%%%%%%%%%%%%%%%%%%
\subsection{Basics from Group Theory and Their Presentations}
\label{s:Basics}
This subsection briefly recalls some constructions from group theory and
their properties
which will heavily be used later on. For a more detailed exposition
as well as proofs of the cited results we refer
to the two textbooks \cite{Lyndon,Rotman}. Our notational emphasis 
for each construction and claim
lies on the particular group \emph{presentation} under consideration
--- for two reasons:
First and as opposed to the
discrete case\footnotemark[\ref{myFn}], different presentations
of the same group may heavily affect its effectivity
properties (Example~\ref{x:Rationals}). 
And second, sometimes there does not seem to be a
`natural' choice for a presentation
(Remark~\ref{r:Intersect},
Footnote~\ref{f:Union}).

Here, no (e.g. effectivity)
assumptions are made concerning the set of generators nor
relations presenting a group.
To start with and just for the records, let us briefly extend
the standard notions of a subgroup and a homomorphism
to the setting of \emph{presented} groups:
\begin{definition} \label{d:Presented}
A \emph{subgroup} $U$ of the presented group $G=\langle X|R\rangle$
is a tuple $(V,S)$ with $V\subseteq\langle X\rangle$ 
and $S=R\cap\langle V\rangle$.
This will be denoted by $U=\langle V|R_V\rangle$
or, more relaxed, $U=\langle V|R\rangle$.
\\
A \emph{realization} of a homomorphism
$\psi:G\to H$ between presented groups
$G=\langle X|R\rangle$ and $H=\langle Y|S\rangle$ 
is a mapping $\psi':X\to\langle Y\rangle$
whose unique extension to a homomorphism on
$\langle X\rangle$ maps $R$-cosets to $S$-cosets,
that is, makes Equation~(\ref{e:Embedding}) commute.
\\
A realization of an isomorphism $\phi$
is a realization of $\phi$ as a homomorphism.
\end{definition}
In the above notation,
$\langle\psi'(X)\big|S\rangle$
is a presentation of the subgroup $\psi(G)$ of $H$.
For an embedding $\psi$,
$G$ is classically isomorphic to $\psi(G)$;
Lemma~\ref{l:Embedding} below contains a 
computable variation of this fact.
\begin{remark} \label{r:Intersect}
The intersection $A\cap B$ of two subgroups $A,B$ of $G$
is again a subgroup of $G$. For presented sub-groups
$A=\langle U|R\rangle$ and $B=\langle V|R\rangle$ of
$G=\langle X|R\rangle$ however, $\langle U\cap V|R\rangle$
is in general \emph{not} a presentation of $A\cap B$.
\end{remark}

\begin{definition}[Free Product] \label{d:Free}
Consider two presented groups $G=\langle X|R\rangle$ 
and $H=\langle Y|S\rangle$ with disjoint generators
$X\cap Y=\emptyset$ ---
e.g. by proceeding to $X':=X\times\{1\}, Y':=Y\times\{2\}$,
$R':=R\times\{1\}$, $S':=S\times\{2\}$.
The \emph{free product} of $G$ and $H$ is the presented group
\[ G\freeprod H \quad:=\quad
  \big\langle X\cup Y \;\big|\; R\cup S\big\rangle 
\enspace . \]
Similarly for the free product
$\displaystyle\Freeprod_{i\in I} G_i$ with
$G_i=\langle X_i|R_i\rangle$, $i$ ranging
over arbitary index set $I$.%
\end{definition}
In many situations one wants to identify certain elements
of a free product of groups. These are provided by
two basic constructions:
\emph{amalgamation} and \emph{Higman-Neumann-Neumann}
(or shortly HNN) extension, see \cite{Higman-Neumann,Lyndon,Rotman}.
The intuition behind the latter is nicely illustrated, e.g., 
in \mycite{Figure~11.9}{Rotman}.

\begin{definition}[Amalgamation] \label{d:Amalgam}
Let $G=\langle X|R\rangle$, $H=\langle Y|S\rangle$ with $X\cap Y=\emptyset$.
Let $A=\langle V|R\rangle$ and $B=\langle W|S\rangle$ be
respective subgroups and $\phi':\langle V\rangle\to\langle W\rangle$
realization of an isomorphism $\phi:A\to B$. 
%in the sense of Definition~\ref{d:Presented}.
The free product of $G$ and $H$ \emph{amalgamating} the subgroups $A$
and $B$ via $\phi$ is the presented group
\begin{equation} \label{e:Amalgam}
\langle G*H\;|\;\phi(a)=a\forall a\in A\rangle
\quad:=\quad 
\big\langle X\cup Y\;|\;R\,\cup\, S\,\cup\,\{\phi'(\bar v)\bar v^{-1}:
\bar v\in V\}\big\rangle 
\enspace .
\end{equation}
\end{definition}

\begin{definition}[HNN Extension] \label{d:HNN}
Let $G=\langle X|R\rangle$, $A=\langle V|R\rangle,B=\langle W|R\rangle$
subgroups of $G$, and 
$\phi'$ realization of an isomorphism between $A$ and $B$. The 
\emph{Higman-Neumann-Neumann} (HNN) extension
of $G$ relative to $A,B$ and $\phi$ is the presented group
\[ \langle G;t\;|\; ta=\phi(a)t\forall a\in A\rangle
\quad:=\quad
\big\langle X\cup\{t\}\;|\;R\,\cup\,
\{\phi'(\bar v)t\bar v^{-1}t^{-1}:\bar v\in V\}\big\rangle 
\enspace . \]
$G$ is the \emph{base} of the HNN extension, $t\not\in X$ 
is a new generator called the  \emph{stable letter}, and
$A$ and $B$ are the \emph{associated subgroups} of the extension.

Similarly for the HNN extension 
$\langle G; (t_i)_{i\in I} \,|\,
  t_ia=\phi_i(a)t_i\forall a\in A_i\forall i\in I\rangle$
with respect to a family of isomorphisms $\phi_i: A_i \to B_i$ 
and subgroups $A_i,B_i\subseteq G$, $i \in I$.
\end{definition}
Both HNN extensions and free products with amalgamation 
admit simple and intuitive characterizations for
a word to be, in the resulting group, equivalent to 1.
These results are connected to some very famous names
in group theory. Proofs can be found, e.g., in
\mycite{Chapter~IV}{Lyndon} or \mycite{Chapter~11}{Rotman}.

\begin{fact}[Higman-Neumann-Neumann] \label{f:HNN}
%\comment{Britton setzt $n\geq1$ voraus}
Let $G^*:=\langle G;t|ta=\phi(a)t\forall a\in A\rangle$ denote
a HNN extension of $G$. Then,
identity $g\mapsto g$ is an embedding of $G$ into $G^*$.
\qed\end{fact}

\begin{fact}[Britton's Lemma] \label{f:Britton}
Let $G^* :=\langle G;t| ta=\phi(a)t\forall a\in A\rangle$ 
be an HNN extension of $G$.
Consider a sequence $(g_0,t^{\epsilon_1},g_1,\ldots, t^{\epsilon_n},g_n)$
with $n\in\IN$, $g_i \in G$, $\epsilon_i \in \{-1,1\}$.
If it contains no consecutive subsequence $(t^{-1},g_i,t)$
with $g_i \in A$ nor $(t,g_j,t^{-1})$ with $g_j \in B$,
then it holds
$g_0\cdot t^{\epsilon_1}\cdot g_1\cdots t^{\epsilon_n}\cdot g_n\neq 1$ in $G^*$.
\qed\end{fact}

\begin{fact}[Normal Form] \label{f:Normalform}
Let $P:=\langle G*H|\phi(a)=a\forall A\rangle$
denote a free product with amalgamation. Consider
$c_1,\ldots,c_n\in G*H$, $n\in\IN$, such that
\begin{itemize}
\item[--] each $c_i$ is either in $G$ or in $H$;
\item[--] consecutive $c_i,c_{i+1}$ come from different factors;
\item[--] if $n >1,$ then no $c_i$ is in $A$ nor $B$;
\item[--] if $n=1,$ then $c_1 \neq 1.$
\end{itemize}
Then, $c_1\cdots c_n\neq 1$ in $P$.
\qed\end{fact}

%%%%%%%%%%%%%%%%%%%%%%%%%%%%%%%%%%%%%%%%%%%%%%%%%%%%%%
\subsection{First Effectivity Considerations} \label{s:Effectivity}
Regarding finitely generated groups, the cardinalities of the sets 
of generators (that is their \emph{ranks}) add under free products
\mycite{Corollary~\textsection IV.1.9}{Lyndon}. 
Consequently, they can straight forwardly be bounded under 
both HNN extensions and free products with amalgamation. 
Similarly for real groups, we have easy control over 
the \emph{dimension} $N$ 
of set of generators according to Definition~\ref{d:Groups1}:
\begin{observation} \label{o:Dimension}
For groups $G_i=\langle X_i|R_i\rangle$ 
with $X_i\subseteq\IR^N$ for all $i\in I\subseteq\IR$,
the free product%
\[ \Freeprod_{i\in I} G_i \quad=\quad %$ is (effectively isomorphic to)
\big\langle \bigcup\nolimits_{i\in I}
({X\times\{i\}}) \;\big|\; \bigcup\nolimits_{i\in I} (R\times\{i\}) \big\rangle
\]
is of dimension at most $N+1$.
In the countable case $I\subseteq\IN$,
the dimension can even be achieved to not
grow at all: by means of a bicomputable bijection
$\IR\times\IN\to\IR$ like
$(x,n)\mapsto\langle\lfloor x\rfloor,n\rangle+(x-\lfloor x\rfloor)$.%
\\
Similarly for free products with amalgamation
and for HNN extensions.
\end{observation}
Moreover, free
%} Free 
products, HNN extensions, and amalgamations
of \SAGEP groups are,
under reasonable presumptions, 
again \SAGEP:
\begin{lemma} \label{l:Hereditary}
\begin{enumerate}
\item[a)]
 Let $G_i=\langle X_i|R_i\rangle$ for all $i\in I\subseteq\IN$.
 If $I$ is finite and each $G_i$
 \SAGEP,
 then so is $\Freeprod_{i\in I} G_i$.
\\
 Same for $I=\IN$, provided that $G_i$ is 
 \SAGEP
 \emph{uniformly in $i$}.
\item[b)]
 Let $G=\langle X|R\rangle$ and consider the HNN extension 
 $G^*:=\langle G; (t_i)_{i\in I} \,|\,
  t_ia=\phi_i(a)t_i\forall a\in A_i\forall i\in I\rangle$
  with respect to a family of isomorphisms $\phi_i: A_i \to B_i$ 
  between subgroups $A_i=\langle V_i|R\rangle,B_i=\langle W_i|R\rangle$
  for $V_i,W_i\subseteq\langle X\rangle$, $i \in I$.
\\
 Suppose that $I$ is finite, each $G_i$ is 
 \SAEP,
 $V_i\subseteq\ri$ is semi-/decidable,
 and finally each $\phi_i$ is effective as a homomorphism;
 then $G^*$ is \SAEP as well.
\\
 Same for $I=\IN$, provided that the $V_i$ are \emph{uniformly} 
 semi-/decidable
 and effectivity of the $\phi_i$ holds \emph{uniformly}.
\item[c)]
 Let $G=\langle X|R\rangle$ and $H=\langle Y|S\rangle$; let
 $A=\langle V|R\rangle\subseteq G$ and $B=\langle W|S\rangle\subseteq H$ 
 be subgroups
 with $V\subseteq\langle X\rangle$, $W\subseteq\langle Y\rangle$,
 $V\subseteq\ri$ semi-/decidable,
 and $\phi:A\to B$ an isomorphism and effective homomorphism.
 Then, their free product with amalgamation (\ref{e:Amalgam})
 is \SAEP whenever $G$ and $H$ are.
\end{enumerate}
\end{lemma}
\begin{remark} \label{r:Uniform}
\emph{Uniform} (semi-)decidability of a family $V_i\subseteq\ri$
of course means that every $V_i$ is (semi-)decidable not
only by a corresponding \BCSS-machine $\IM_i$, but all 
$V_i$ by one common machine $\IM$;
similarly for \emph{uniform} computability
of a family of mappings.
By virtue of (the proof of)
\mycite{Theorem~2.4}{Cucker}, a both necessary
and sufficient condition for such uniformity is that
the real constants employed by the $\IM_i$ 
can be chosen to all belong to one common finite field
extension $\IQ(c_1,\ldots,c_k)$ over the rationals.
\qed\end{remark}
Recall (Observation~\ref{o:Embedding}) that a homomorphism
between finitely generated groups is automatically effective
and, if injective, has decidable range and effective inverse.
For real groups however, in order to make sense out of
the prerequisites in Lemma~\ref{l:Hereditary}b+c),
we explicitly have to specify the following
\begin{definition} \label{d:Embedding}
 An homomorphism
 $\psi:\langle X|R\rangle\to\langle Y|S\rangle$
 of presented real groups %with $X,Y\subseteq\ri$ 
 is called an \emph{effective homomorphism} if it
 admits a \BCSS-computable realization
 $\psi':X\to\langle Y\rangle$
 in the sense of Definition~\ref{d:Presented}.

 For $\psi$ to be called an \emph{effective embedding},
 it must not only be an effective homomorphism and injective; 
 but $\psi'$ is also required to be injective
 and have decidable image $\psi'(X)$
 plus a \BCSS-computable inverse
 $\chi':\psi'(X)\subseteq\langle Y\rangle\to X$.
% whose unique extension to a homomorphism
% on $\psi'(\langle X\rangle)$ maps $S$-cosets to $R$-cosets.
% GILT AUTOMATISCH, DA PSI+PSI' BEIDE INJEKTIV!
\end{definition}
Effective embeddings arise in Lemmas~\ref{l:Embedding}
and \ref{l:Benign}.
For an injective effective homomorphism $\psi$ as
in Lemma~\ref{l:Hereditary}c) on the
other hand, a realization need not be injective;
for instance, $\psi'$ might map two
equivalent (w.r.t. the relations $R$)
yet distinct words to the same image word.
\begin{proof}[Lemma~\ref{l:Hereditary}]
\begin{enumerate}
\item[a)]
If $X_i$ is decidable for each $i\in I$, $I$ finite,
then so is $\bigcup_{i\in I} (X_i\times\{i\})$;
same for semi-decidable/decidable $R_i$.
\emph{Uniform} (semi-)decidability of each $X_i$ means
exactly that $\bigcup_{i\in\IN} (X_i\times\{i\})$ is 
(semi-)decidable.
\item[b)]
The set of generators of the HNN extension is
decidable as in a). The additional relations
$\{\phi'(\bar v)t\bar v^{-1}t^{-1}:\bar v\in V\}$ are semi-/decidable since, 
by presumption, $V$ is and $\phi':\langle V\rangle\to\langle W\rangle$ 
is computable. Uniformity enters as in a).
\item[c)]
Similarly.
\qed\end{enumerate}\end{proof}

\begin{lemma} \label{l:Embedding}
  Let $\psi:G=\langle X|R\rangle\to\langle Y|S\rangle=K$
  denote an effectively realizable embedding.
\begin{enumerate}
\item[a)]
  There is an effectively realizable embedding
  $\chi:\psi(G)\to G$ (i.e. we have an effective isomorphism).
\item[b)]
  If $V\subseteq\langle X\rangle$ is decidable,
  then the restriction $\psi|_H$ to $H=\langle V|R\rangle\subseteq G$
  is an effectively realizable embedding again.
\item[c)]
  If $G$ is \SAG and $K$ \SAP
  then $\psi(G)$ is \SAP as well.
\end{enumerate}
\end{lemma}
\begin{proof}
\begin{enumerate}
\item[a)]
Let $\psi':X\to\langle Y\rangle$ denote the effective realization
of $\psi$ with inverse $\chi'$ according to Definition~\ref{d:Embedding}.
The unique extension of $\psi'$ to a homomorphism
has image $\psi'(\langle X\rangle)=\langle\psi'(X)\rangle$.
Similar to Lemma~\ref{l:DecidableSpan}
we can decide, given $\bar w\in\langle Y\rangle$,
whether $\bar w\in\psi'(\langle X\rangle)$.
Moreover if so, we obtain
a partition $\bar w=(\bar v_1,\ldots,\bar v_\ell)$
with $\bar v_i\in\psi'(X)$.
Then calculating $x_i:=\chi'(\bar v_i)\in X$ yields
a computable extension of $\chi'$ to a
homomorphism on $\psi'(\langle X\rangle)$
which satisfies injectivity, has decidable
image and $\psi'$ as inverse.
Moreover $\chi'$ maps $S$-cosets
to $R$-cosets:
Take $\bar v_1,\bar v_2\in\psi'(\langle X\rangle)$
with $\bar v_1/S=\bar v_2/S$;
then $\bar u_i:=\chi'(\bar v_i)$
have $\bar v_i=\psi'(\bar u_i)$
and thus, since $\psi'$ makes
Equation~(\ref{e:Embedding}) commute by presumption,
$\bar v_1/S=\psi(\bar u_1/R)=\psi(\bar u_2/R)=\bar v_2/S$;
now injectivity of $\psi$ implies
$\bar u_1/R=\bar u_2/R$.
\item[b)]
The range $\psi'(V)$ of the restriction $\psi'|_V$
coincides with $\chi'^{-1}(V)\cap\langle\psi'(X)\rangle$.
The first term is decidable since
$\chi'$ is computable and $V$ decidable;
the second term is decidable by
Definition~\ref{d:Embedding} and Lemma~\ref{l:DecidableSpan}.
\item[c)]
Becomes clear by staring at $\psi(G)=\langle\psi'(X)|S\rangle$.
\qed\end{enumerate}\end{proof}
%%%%%%%%%%%%%%%%%%%%%%%%%%%%%%%%%%
\subsection{Benign Embeddings} \label{s:Benign}
The requirement in Lemma~\ref{l:Hereditary}b+c) that
the subgroup(s) $A$ be recursively enumerable or
even decidable, is of course central but 
unfortunately violated in many cases.
For instance, a subgroup of a finitely presented group 
in general need not even be finitely generated:
Consider, e.g., the \emph{commutator} 
$[G,G]:=\langle\{u v u^{-1} v^{-1}:u,v\in G\}\rangle$
of the free group $G=\langle\{a,b\}\rangle$ and compare
Remark on p.177 of \cite{Lyndon}. 
Similarly the \SAP real group $(\IR,+)$ 
has a subgroup 
(Example~\ref{x:Rationals}a) which is not \SAG.
Nevertheless, both can obviously be effectively
embedded into a, respectively, finitely presented
and an \SAP group.
This suggests the notion of \emph{benign} subgroups,
in the classical case (below, Item~a) introduced in \cite{Higman}.
Recall that there, effectivity of an embedding drops off automatically.
\begin{definition} \label{d:Benign}
\begin{enumerate}
\item[a)]
 Let $X$ be finite, $V\subseteq\langle X\rangle$.
 The subgroup $A=\langle V|R\rangle$ of 
 $G=\langle X|R\rangle$ is (classically) \emph{benign in $G$}
  if the HNN extension
 $\langle X;t\,|\, ta=at\forall a\in A\rangle$
 can be embedded into some finitely presented group
 $K=\langle Y|S\rangle$.
\item[b)]
 Let $X\subseteq\ri$, $V\subseteq\langle X\rangle$.
%\comment{Im Hinblick auf Lemma~\ref{l:Benign}a)
%ausdr\"{u}cklich NICHT fordern, $G$ m\"{u}sse SAG sei!}
 The subgroup $A=\langle V|R\rangle$ of 
 $G=\langle X|R\rangle$ is \emph{effectively benign 
 in $G$} if the HNN extension
 $\langle G;t\,|\, ta=at\forall a\in A\rangle$
 admits an effective embedding
 into some \SAP group $K=\langle Y|S\rangle$.
\item[c)]
 Let $I\subseteq\IN$. A family 
 $(A_i)_{_{i\in I}}$ of subgroups of $G$ 
 is \emph{uniformly effectively benign in $G$}
 if, in the sense of Remark~\ref{r:Uniform}, 
 there are groups $K_i$ uniformly \SAP and
 uniformly effective embeddings 
 $\phi_i:\langle G;t_i|t_ia_i=a_it_i\forall a_i\in A_i\rangle\to K_i$.
\end{enumerate}\end{definition}
%By Lemma~\ref{l:Hereditary}, the HNN extension in b)
%is effectively generated; hence the Definition~\ref{d:Embedding}
%of an effective embedding indeed applies.
The benefit of benignity is revealed in the following
\begin{remark} \label{r:Benign}
In the notation of Definition~\ref{d:Benign}b),
if $A$ is effectively benign in $G$
then the word problem for $A$
is reducible to that for $K$:
Fact~\ref{f:HNN}.
%Formally: given a word $\bar w\in\langle V\rangle$...
\\
Moreover in this case, 
the \emph{membership problem} for $A$ in $G$
--- that is the question whether 
given $\bar x\in\langle X\rangle$
is equivalent (w.r.t. $R$) to an
element of $A$ --- is also reducible
to the word problem for $K$:
According to Fact~\ref{f:Britton},
$a:=\bar x/R$ satisfies
\quad $t\cdot a\cdot t^{-1}\cdot a^{-1}=1
\Leftrightarrow a\in A$.%
\qed\end{remark}
We now collect some fundamental properties
frequently used later on. 
They extend corresponding results from the
finite framework. Specifically,
Lemma~\ref{l:Benign}b) generalizes 
\cite[\textsc{Lemma~\textsection IV.7.7}(i)]{Lyndon}
and Claims~d+e) generalize
\cite[\textsc{Lemma~\textsection IV.7.7}(ii)]{Lyndon}.
\begin{lemma} \label{l:Benign}
\begin{enumerate}
\item[a)]
  Let $A=\langle V|R\rangle\subseteq H=\langle W|R\rangle\subseteq G=\langle X|R\rangle$
  denote a chain of sub-/groups
  with $V\subseteq\langle W\rangle$ and $W\subseteq\langle X\rangle$.
  If $W$ is decidable and $A$ effectively benign in $G$,
  then it is also effectively benign in $H$.
\item[b)]
  If $G=\langle X|R\rangle$ is \SAP
  and subgroup $A=\langle V|R\rangle$
  has decidable generators $V\subseteq\langle X\rangle$,
  then $A$ is effectively benign in $G$.
\item[c)]
  If $A$ is effectively benign in $G$
  and $\phi:G\to H$ an effective embedding,
  then $\phi(A)$ is effectively benign in $\phi(G)$.
\item[d)]
  Let $A$ and $B$
%scheint unnoetige Vor.:  with decidable $V,W\subseteq\langle X\rangle$
  be effectively benign in \SAP $G$.
  Then $A\cap B$ admits a presentation
  effectively benign in $G$.
\item[e)]
  Let $A$, $B$, $G$ as in d);
  then $\langle A\cup B\rangle_G$ admits a 
  presentation\footnote{possibly different from
  \label{f:Union} $\langle V\cup W|R\rangle$}
  effectively benign in $G$.
\item[f)]
  Let $(A_i)_{i\in I}$ be uniformly effectively benign
  in $G$ (Definition~\ref{d:Benign}c).
  Then $\langle\bigcup_{i\in I} A_i\rangle$ admits a
  presentation effectively benign in $G$.
\end{enumerate}
\noindent 
The above claims hold uniformly in that the corresponding 
effective embeddings do not introduce new real constants.
\end{lemma}
\begin{proof}
\begin{enumerate}
\item[a)] 
  Let $\psi$ denote an effectively realizable
  embedding of the HNN extension
  $\langle X;t|ta=\phi(a)t\forall a\in A\rangle$ into some
  \SAP $K=\langle Y|S\rangle$.
  Since $W\cup\{t\}$ is decidable,
  Lemma~\ref{l:Embedding}b)  asserts
  the restriction of $\psi$ to yield
  an effective embedding of the HNN extension
  $\langle W;t|ta=\phi(a)t\forall a\in A\rangle$
  into $K$.
\item[b)] 
  The identity being an effectively realizable embedding
  ($X$ is decidable, now apply Lemma~\ref{l:Embedding}b),
  it suffices to observe that the HNN extension
  \[ K \quad:=\quad \langle G;t\,|\, at=ta\forall a\in A\rangle
  \quad=\quad \langle X;t\,|\, R\cup\{\bar vt=t\bar v\forall \bar v\in V\}\rangle \]
  is \SAP itself. Indeed, $X$, $R$, and the additional relations
  parametrized by $V$ are decidable by presumption.
\item[c)] 
  The presented HNN extension under consideration,
  \begin{equation} \label{e:Temp1}
  \langle\phi'(X);s\,|\,\phi'(\bar v)s=s\phi'(\bar v)\forall\bar v\in V\rangle \enspace ,
  \end{equation}
  is the image under $\phi$ of $\langle G;t|at=ta\forall a\in A\rangle$
  by extending $\phi'(t):=s$. The latter HNN extension
  by presumption embeds into some (finite-dim.) \SAP $K$ via some effective $\psi$.
  According to Lemma~\ref{l:Embedding}a), $\phi$ admits an
  effective inverse. Hence the composition
  $\psi\circ\phi^{-1}$ consitutes the desired effective embedding
  of (\ref{e:Temp1}) into $K$.
\item[d)]
  By assumption there exist two \SAP groups $K=\langle Y|S\rangle$
  and $L =\langle Z|T\rangle$ together with realizations 
  $\phi':X\cup\{r\}\to\langle Y\rangle$,
  $\psi':X\cup\{r\}\to\langle Z\rangle$
  of effective embeddings 
%  $\phi:\langle X;r\,|\,R\cup\{\bar vr=r\bar v:\bar v\in V\}\rangle\to\langle Y|S\rangle$
%  and
%  $\psi:\langle X;r\,|\,R\cup\{\bar wr=r\bar w:\bar w\in W\}\rangle\to\langle Z|T\rangle$
%  of HNN extensions $\langle G;r|ar=ra\forall a\in A\rangle$ and 
%  $\langle G;r|br=rb\forall b\in B\rangle$ into $K$ and $L$, respectively.
  \begin{alignat*}{5}
  \phi\;:\;&G_A \;:=\; \langle G;r|ar=ra\forall a\in A\rangle& %\label{e:BenignA}
       &\;=\;\langle X;r\,|\,R\cup\{\:\bar vr=r\bar v&:&\bar v\in V&\}\rangle
          &\;\to\;&K\;=\;\langle Y|S\rangle&  \\
%  \text{and }\quad 
  \psi\;:\;&G_B\;:=\;\langle G;r|br=rb\forall b\in B\rangle&
       &\;=\;\langle X;r\,|\,R\cup\{\bar wr=r\bar w&:&\bar w\in W&\}\rangle
          &\;\to\;&L\;=\;\langle Z|T\rangle& \label{e:BenignB}
  \enspace . \end{alignat*}
  We shall realize an embedding
  $\chi$ of the HNN extension $G_C:=\langle G;r|cr=rc\forall c\in C\rangle$
  into an \SAP group for the presentation\footnote{Notice\label{f:Intersect}
  the arbitrarily broken symmetry between the groups/embeddings 
  $(A,\phi)$ and $(B,\psi)$ involved.}
  $C:=\big\langle\{\bar w\in\langle W\rangle:\bar w/R\in A\}\,\big|\,R\big\rangle$
  for $A\cap B$.
  To this end observe that $\phi(G)=\langle\phi'(X)|S\rangle$
  and $\psi(G)=\langle\psi'(X)|T\rangle$ are subgroups of
  $K$ and $L$, respectively, and isomorphic due to 
  Fact~\ref{f:HNN} with isomorphism\footnotemark[\ref{f:Intersect}]
  $\phi\circ\psi^{-1}:\psi(G)\to\phi(G)$ realized by
  $\phi'\circ\psi'^{-1}$ according to
  Lemma~\ref{l:Embedding}. 
  Definition~\ref{d:Amalgam} is thus
  applicable and we are entitled to consider 
  the free group with amalgamation
  \begin{eqnarray} \label{e:P}
  P &:=& \big\langle K*L \,\big|\, 
    \phi\big(\psi^{-1}(\ell)\big)=\ell\forall \ell\in\psi(G)\big\rangle 
  \\ &=&
  \big\langle Y\cup Z\,\big|\,S\cup T\cup\{
   \phi'\big(\psi'^{-1}(\bar z)\big)=\bar z:\bar z\in\psi'(X)\}
   \big\rangle \enspace . \nonumber
  \end{eqnarray}
  $P$ is \SAP because of Lemma~\ref{l:Hereditary}c).
  Moreover $\phi(G)=\psi(G)$ in $P$ according to (\ref{e:P}).
  Also, $s:=\phi'(r)$ commutes exactly with $\phi(A)$
  and $t:=\psi'(r)$ exactly with $\psi(B)$,
  so $s\cdot t$ commutes exactly with $\phi(A)\cap\psi(B)$.
  Therefore,  $\chi':X\cup\{r\}\to\langle Y\cup Z\rangle$,
  $x\mapsto\psi'(x)$, $r\mapsto s\cdot t$
  respects cosets in the sense of Equation~(\ref{e:Embedding}) 
  and thus realizes an embedding 
  $\chi:\langle G;r|cr=rc\forall c\in C\rangle\to P$
  as desired.
\item[e)] 
  With notations as in d), it holds
  \begin{eqnarray*}
  \psi(\langle A\cup B\rangle_G) \;\;=\;\; \phi(\langle A\cup B\rangle_G)
  &\;\;=\;\;& \langle\phi(A)\cup\phi(B)\big\rangle_P \;\;=\;\;
      \langle\phi(A)\cup\psi(B)\big\rangle_P  \\
   &\;\;=\;\;& \langle \phi(r\cdot G\cdot r^{-1}) \;\cup\; \psi(r\cdot G\cdot r^{-1})
       \rangle_P \;\:\cap\:\; \phi(G) \enspace ;
  \end{eqnarray*}
  the first line because $\phi$ and $\psi$
  are injective homomorphisms coinciding on $G$;
  the second because $A$ and only $A$ commutes
  with $r$ in $G_A$ due to Britton's Lemma (Fact~\ref{f:Britton}),
  similarly for $B$ in $G_B$.
  Now $\phi(G)$ is \SAP due to Lemma~\ref{l:Embedding}c)
  and thus effectively benign in $P$ by Claim~b).
  Similarly,
  $\langle \phi(r\cdot G\cdot r^{-1})\cup\psi(r\cdot G\cdot r^{-1})\rangle_P$
  has decidable generators and is thus effectively
  benign in $P$ as well.
  Claim~d) now asserts effective benignty of
  $\phi(\langle A\cup B\rangle)$ in $P$;
  and therefore also in $\phi(G)\subseteq P$
  according to Claim~a) combined with Lemma~\ref{l:Embedding}c).
  Claim~c) combined with Lemma~\ref{l:Embedding}a)
  finally yields effective benignty of $\langle A\cup B\rangle$
  in $G$.
\item[f)]
  Let $(\phi_i')_{i\in I}$ denote the uniformly computable realizations
  of embeddings $\phi_i:G_i:=\langle G;r|ar=ra\forall a\in A_i\rangle\to K_i$.
  Fix $j\in I$. Similar to Equation~\ref{e:P}) and the proof of e), we have
  \begin{gather*}
  \phi_j\Big(\big\langle\Freeprod_{i\in I} A_i\big\rangle_G\Big)
  \;=\; \Big\langle\bigcup_{i\in I}\phi_i\big(r^{}\cdot G\cdot r^{-1}\big)
   \Big\rangle_P \;\cap\;\; \phi_j\big(G\big), \\
    P \;:=\;
  \Big\langle\Freeprod_{i\in I} K_i \,\Big|\,
     \phi_i^{}\big(\phi_j^{-1}(\ell)\big)=\ell\,
       \forall\ell\in\phi_j(G)\,\forall i\in I\Big\rangle
  \end{gather*}
  where (by uniformity, see Lemma~\ref{l:Hereditary}) 
  $P$ and $\phi_j(G)$ are \SAP,
  and $\big\langle\bigcup_{i\in I}\phi_i(r\cdot G\cdot r^{-1})
   \big\rangle_P$
  has decidable generators of bounded dimension,
  compare Observation~\ref{o:Dimension}.
\qed\end{enumerate}\end{proof}
We are now ready to indulge into the main part of the proof.

%%%%%%%%%%%%%%%%%%%%%%%%%%%%%%%%%%%%%%%%%%%%%
\subsection{Dealing with a single path set} \label{s:Single}
Consider the real halting problem $\IH\subseteq\ri$
together with an appropriate 
\BCSS machine $\IM$ which accepts exactly inputs $\bar r$ belonging to $\IH$ and
stalls for all others.
The accepting paths of $\IM$ admit an effective enumeration
$(\gamma_n)$, $n \in \N$. Here, each path $\gamma_n$
is described by
a finite sequence (of length $D=D(\gamma_n)\in\IN$, say)
of primitive arithmetic operations, assignments, 
and comparisons performed along it. Each such path $\gamma$ gives rise to
the (possibly empty) set $\IA_{\gamma}\subseteq\IR^{d}$, $d=d(\gamma)\in\IN$,
of inputs $\bar r\in\IR^d$ on which $\IM$ follows exactly this path.
Both functions $n\mapsto d(\gamma_n)$ and $n\mapsto D(\gamma_n)$ are
computable. 

A computational path $\gamma$ and input $(r_1,\ldots,r_d)$ following it,
gives rise to a sequence $r_{d+1},\ldots,r_{D}\in\IR$
of intermediate results, each one being the result from a composition of
at most two previous ones. For instance, $r_i=r_j\pm r_k$ with $d<i\leq D$ and
$1\leq j,k<i$; or $r_i=\alpha$ for some machine constant $\alpha\in\IR$ of $\IM$;
branches take the form ``$r_i \geq 0?$''.%; and by convention, $r_D=1$.
The advantage of this description of $\gamma$ as a set 
$\IB_\gamma\subseteq\IR^D$ of $(r_1,\ldots,r_d,r_{d+1},\ldots,r_D)$
is that each intermediate result $r_i$
may be accessed several times but gets assigned only once.

In view of Remark~\ref{r:Benign}, 
our goal is to write %(first $\IB_n$ and then) 
$\IA_{\gamma_n}$ as a subgroup $U_{\gamma_n}$
effectively benign such that membership to $\IA_{\gamma_n}$
is reducible to that of $U_{\gamma_n}$;
with the additional constraint that all
constructions work \emph{uniformly} in $n$ ---
in fact using only constants already 
present in $\IM$;
compare Remark~\ref{r:Uniform} and see 
Footnote~\ref{f:Constants}.
However for notational convenience, 
$n$ (and thus also $\gamma,d,D$) will be kept fixed 
and occasionally omitted throughout this subsection.
They reappear in Section~\ref{s:Final} when the subgroups 
$U_{\gamma_n}$, $n\in\IN$, are finally glued together.

\begin{definition} \label{d:Single}
Let 
\[ X\;:=\;\{x_{(i,s)}:s\in\IR,i\in\IN\}\cup\{y\}\;\cong\;(\IR\times\IN)\cup\{\infty\}, 
\quad G\;:=\;\langle X\rangle \]
denote a free group with subgroups
\[ H_{\leq d}:=\langle \{y,x_{(i,s)}:s\in\IR,i\leq d\}\rangle
\quad\text{and}\quad
H_{>d}:=\langle x_{(i,s)}:s\in\IR,i>d\rangle \enspace . \]
Furthermore consider the subgroups
\[ U_\gamma\;:=\;\langle\bar w_{\bar r}:\bar r\in\IA_\gamma\rangle
\quad\text{and}\quad
V_\gamma\;\;:=\;\;\langle\bar w_{\bar s}:\bar s\in\IB_\gamma\rangle \]
with the abbreviation
$\bar w_{(r_1,\ldots,r_k)}:=x^{-1}_{(k,r_k)}\cdots x^{-1}_{(1,r_1)}\cdot y
\cdot x^{}_{(1,r_1)}\cdots x^{}_{(k,r_k)}$
\ for $r_1,\ldots,r_k\in\IR$.
\end{definition}
The reason for the complicated definition of $\bar w$
(as opposed to using, e.g., $\bar v_{\bar r}:=x_{(1,r_1)}\cdots x_{(k,r_k)}$)
lies in the following
\begin{fact} \label{f:Nielsen}
The words $\bar w_{\bar r}$, $\bar r\in\IA_\gamma$,
are \emph{Nielsen-reduced}---compare {\rm\cite[p.223]{Lyndon}}---and 
thus freely generate $U_\gamma$
{\rm\mycite{Proposition~\textsection I.2.5}{Lyndon}}.
In particular, $\bar w_{\bar r}\in U_\gamma$ iff $\bar r\in\IA_{\gamma}$.
\end{fact}

\begin{theorem} \label{t:Single}
$U_\gamma$ is (or rather, has a presentation) effectively benign in \SAP $G$.
\end{theorem}
Let $(o_{d+1},\ldots,o_D)$ denote the arithmetic operations, 
assignments, and branched tests performed on the path $\gamma$;
cf. left column of Figure~\ref{f:OpGroups}.
For each such $o$, define a subgroup $W_o$ of $G$ as in
the middle column of Figure~\ref{f:OpGroups}.
Since the generators involved are free, we have
%(Fact~\ref{f:Normalform})
\begin{lemma} It holds \label{l:Single}
\vspace*{-2ex}\[
  \displaystyle V_\gamma\;=\;\bigcap\nolimits_{i=d+1}^D W_{o_i} \vspace*{-1ex} \]
and\vspace*{-1ex}%
%\smallskip
%\item[b)]
\[ \displaystyle U_\gamma\;=\;\langle V_\gamma\:\cup\:H_{>d}\rangle
    \;\cap\; H_{\leq d} \vspace*{-1ex}\]
%\end{enumerate}
\end{lemma}
\begin{proof}%[Lemma~\ref{l:Single}]
Let us focus on the second claim, the argument for the first one proceeds similarly.
Inclusion ``$U_\gamma\subseteq\langle V_\gamma\cup H_{>d}\rangle\cap H_{\leq d}$'' 
holds since to every 
word $\bar w_{\bar r}\in U_\gamma$,
$(r_1,\ldots,r_d)\in\IA_\gamma$, there corresponds an extension
$\bar w_{\bar s}\in V_\gamma$
with $\bar s=(r_1,\ldots,r_d,r_{d+1},\ldots,r_D)\in\IB_\gamma$;
and the symbols $x_{(i,r)}$ with $i>d$
can be cancelled from $\bar w$ by means of $H_{>d}$, thus transforming into
an element of $H_{\leq d}$.
\\
For the reverse inclusion, observe that the words
$\bar w_{\bar s}\in V_\gamma$
equivalent to a word in $H_{\leq d}$
are exactly those with symbols $x_{(i,r)}$, $i>d$,
removed and with $i\leq d$ unmodified.
\qed\end{proof}
We will now show that the $W_o$ are effectively benign in $G$;
hence Lemma~\ref{l:Benign}d) establishes the same for $V_\gamma$.
Since the respective sets of generators are easily
decidable, Lemma~\ref{l:Benign}b) yields also $H_{\leq d}$
and $H_{>d}$ effectively benign in $G$.
So by Lemma~\ref{l:Benign}e+d), Theorem~\ref{t:Single} follows.

\begin{definition}
Let $C$ denote the infinite (in fact uncountable) HNN extension 
\[ \bigg\langle G \;\;;\;\;
\begin{array}{lr}
a_{(i,t)}&\forall t\in\IR\;\forall i\in\IN \\
m_{(i,t)}&\forall 0\not=t\in\IR\;\forall i\in\IN
\end{array}
\bigg|
\begin{array}{r@{\:=\:}lr}
a_{(i,t)}\cdot g&\,\phi_{(i,t)}(g)\cdot a_{(i,t)}&
\forall g\in G \;\forall (i,t) \\
m_{(i,t)}\cdot g&\psi_{(i,t)}(g)\cdot m_{(i,t)}&
\forall g\in G \;\forall (i,t) \\
\end{array}
\bigg\rangle \]
%\begin{eqnarray*}
%\big\langle B\;;\; a_{(i,t)},m_{(i,t)}\forall t\in\IR\forall i=1,\ldots,D &\big|&
%a_{(i,t)}\cdot x=\phi_{(i,t)}(x)\cdot a_{(i,t)} \;;\\
%&& m_{(i,t)}\cdot x=\psi_{(i,t)}(x)\cdot m_{(i,t)} 
%\;\forall t\in\IR\forall i=1,\ldots,D \big\rangle
%\end{eqnarray*}
with base $G$ and stable letters $a_{(i,t)}$, $m_{(i,t)}$ as above.
Here, $\phi_{(i,t)},\psi_{(i,t)}:G\to G$ denote the 
isomorphisms\footnote{Notice that
$\psi_{(i,t)}$ has $t\not=0$. In fact, we take into account
only \BSS computations which do not multiply with 0. This
is no loss of generality because any multiplication command
may be preceded with a test whether any of the factors 
equals $0$ and, if so, a direct assignment of 0.}
\[ \begin{array}{rlll}
\phi_{(i,t)}:&x_{(i,s)}\mapsto x_{(i,s+t)}, \quad
   & x_{(j,s)}\mapsto x_{(j,s)}, & \quad y\mapsto y \\[0.8ex]
\psi_{(i,t)}:&x_{(i,s)}\mapsto x_{(i,s\cdot t)}, \quad
   & x_{(j,s)}\mapsto x_{(j,s)}, & \quad y\mapsto y
\end{array} \qquad \forall s\in\IR\;\; \forall j\not=i
\enspace . \]
\end{definition}
Intuitively in $C$, commuting a stable letter $a_{(i,t)}$ `causes' 
a real addition in the sense that
$a_{\scriptscriptstyle(i,t)}^{}\cdot x_{(i,s)}\cdot a_{\scriptscriptstyle(i,t)}^{-1}=x_{(i,s+t)}$.
Furthermore, since $a_{(i,t)}$ commutes with all
$x_{(j,s)}$, $j\not=i$, it holds%
\begin{equation} \label{e:Addition}
a_{\scriptscriptstyle(i,t)}^{}\cdot\bar w_{(r_1,\ldots,r_i,\ldots,r_D)}
\cdot a_{\scriptscriptstyle(i,t)}^{-1}
\quad=\quad\bar w_{(r_1,\ldots,r_i+t,\ldots,r_D)}
\enspace ;
\end{equation}
similarly with generators $m_{(i,t)}$ for multiplication.

\stepcounter{footnote}\footnotetext{\label{f:Constants}This is the only
place where real constants occur; however those that do, belong to the 
finitely many already present in $\IM$.}
\begin{lemma} \label{l:Single2}
  For each operation $o$ and its
  corresponding subgroup $L_o$ of $C$ as in Figure~\ref{f:OpGroups},
  it holds $W_o=G\cap L_o$, and
  $W_o$ is effectively benign in $G$.
\end{lemma}
\begin{figure}[htb]
\begin{center}%
\begin{tabular}{l@{\;\;}|@{\;\;}l@{\;\;}|l}
\hspace*{\fill}$o$\hspace*{\fill} & \hspace*{\fill}$W_o\subseteq G$\hspace*{\fill}
& \hspace*{\fill} $L_o\subseteq C$ \hspace*{\fill} %
\\ \hline
``$x_i\leftarrow x_j$'', \hfill $1\leq j<i$%
&
$\langle\bar w_{\bar r}:r_i=r_j\rangle$
&%
\begin{tabular}[t]{rl}%
$\langle\bar w_{\bar 0}\,$&$;\;\;a_{(i,s)}\cdot a_{(j,s)}:s\in\IR\;;$\\
&
$;\;\;a_{(\ell,s)}:s\in\IR,\ell\not=i,j\rangle$
\end{tabular}
\\[3.5ex]
``$x_i\leftarrow\alpha$'',\hfill$\alpha\in\IR$ fixed\footnotemark[\ref{f:Constants}]%
&
$\langle\bar w_{\bar r}:r_i=\alpha\rangle$
&
\,$\langle\bar w_{(0,\ldots,0,\alpha,0,\ldots,0)}\;\;;\;\;a_{(\ell,s)}:s\in\IR,\ell\not=i\rangle$
\\[1ex]
``$x_i\leftarrow x_j+x_k$'',\hfill$1\leq j,k<i$%
&
$\langle\bar w_{\bar r}:r_i=r_j+r_k\rangle$
&
\begin{tabular}[t]{rl}%
$\langle\bar w_{\bar 0}\,$&$;\;\;a_{(\ell,s)}:s\in\IR,\ell\not=i,j,k$\;;\\
&$;\;\;a_{(i,s)}\cdot a_{(k,s)},\;a_{(j,s)}\cdot a_{(k,s)}:s\in\IR\rangle$
\end{tabular}
\\[3.5ex]
``$x_i\leftarrow -x_j$'',\hfill$1\leq j<i$%
&
$\langle\bar w_{\bar r}:r_i=-r_j\rangle$
&
\begin{tabular}[t]{rl}%
$\langle\bar w_{\bar 0}\,$&$;\;\;a_{(\ell,s)}:s\in\IR,\ell\not=i,j$\;;\\
&$;\;\;a_{(i,s)}a_{(j,-s)}:s\in\IR\rangle$
\end{tabular}
\\[3.5ex]
``$x_i\leftarrow x_j\times x_k$'',\hfill$1\leq j,k<i$%
&
$\langle\bar w_{\bar r}:r_i=r_j\cdot r_k\rangle$
&
\begin{tabular}[t]{rl}%
$\langle\bar w_{\bar e_{\{i,j,k\}}}\!\!$&$;\;a_{(\ell,s)}:s\in\IR,\ell\not=i,j,k$\;;\\
&\hspace*{-2.6ex}%
$;\;m_{(i,s)}\cdot m_{(k,s)},\:m_{(j,s)}\cdot m_{(k,s)}:s\in\IR\rangle$
\end{tabular}
\\[3.5ex]
``$x_i\leftarrow 1/x_j$'',\hfill$1\leq j<i$%
&
$\langle\bar w_{\bar r}:r_i=\tfrac{1}{r_j}, r_j\not=0\rangle$
&
\begin{tabular}[t]{rl}%
$\langle\bar w_{\bar e_{\{i,j\}}}\!$&$;\;a_{(\ell,s)}:s\in\IR,\ell\not=i,j$\;;\\
&\hspace*{-1.8ex}%
$;\;m_{(i,s)}\cdot m_{(j,1/s)}:0\not=s\in\IR\rangle$
\end{tabular}
\\[3.5ex]
``$x_j\geq0$'',\hfill$1\leq j<i$%
&
$\langle\bar w_{\bar r}:r_j\geq0\rangle$
&
\begin{tabular}[t]{rl}%
$\langle\bar w_{\bar 0}\,\,$&
$;\;\;a_{(j,s)}:0<s\in\IR$\;;\\
&$;\;\;a_{(\ell,s)}:s\in\IR,\ell\not=j\rangle$
\end{tabular}
\\[3.5ex]
``$x_j<0$'',\hfill$1\leq j<i$%
&
$\langle\bar w_{\bar r}:r_j<0\rangle$
&
\begin{tabular}[t]{rl}%
$\langle\bar w_{(0,\ldots,0,-1,0,\ldots,0)}\,\,$&
$;\;\;m_{(j,s)}:0<s\in\IR$\;;\\
&$;\;\;a_{(\ell,s)}:s\in\IR,\ell\not=j\rangle$
\end{tabular}
\end{tabular}\\[0.5ex]
We abbreviate $\bar 0=(0,\ldots,0)$ and, 
for $I=\{i_1<i_2<\ldots<i_p\}\subseteq\{1,\ldots,D\}$,
$\bar e_{I}:=(0,\ldots,0,\underbrace{1}_{i_1},0,\ldots,0,\underbrace{1}_{i_2},0,
\ldots\ldots\ldots,0,\underbrace{1}_{i_p},0,\ldots,0)$.
\caption{\label{f:OpGroups}Operations and their induced subgroups.}
\end{center}
\end{figure}
\begin{proof}
\begin{description}
\item[$x_i\leftarrow\alpha${\footnotemark[\ref{f:Constants}]}:]
  The inclusion $W_{(x_i\leftarrow\alpha)}\subseteq G\cap L_{(x_i\leftarrow\alpha)}$
  holds because the generators $a_{(j,s)}$ may be used
  according to (\ref{e:Addition})
  to attain, starting from $\bar w_{(0,\ldots,\alpha,\ldots,0)}$,
  any desired value $r_j$ for the symbols $x_{(j,r_j)}$ in
  $\bar w_{\bar r}$, $j\not=i$, while $r_i=\alpha$ cannot be affected.
  Conversely, a representative of an element from $L_{(x_i\leftarrow\alpha)}$
  belonging to $G$ must by Fact~\ref{f:Britton}
  have all stable letters $a_{(\ell,s)}$ removed
  by means of repeated applications of (\ref{e:Addition});
  these leave $r_i=\alpha$ unaffected, thus establishing
  membership to $W_{(x_i\leftarrow\alpha)}$.
\item[$x_i\leftarrow x_j$:]
  Similarly as above, the $a_{(\ell,s)}$ yield,
  starting from $\bar w_{\bar 0}$, $\bar w_{\bar r}$ with any
  value for $r_\ell$, $\ell\not=i,j$; while the
  (by definition of $L_{(x_i\leftarrow x_j)}$ necessarily simultaneous) 
  application of both $a_{(i,s)}$ and $a_{(j,s)}$
  preserves the property ``$r_i=r_j$''.
\item[$x_i\leftarrow x_j+x_k$:]
  Similarly, now preserving ``$r_i=r_j+r_k$''.
\end{description}
\noindent
The other cases proceed analogously
and establish $W_o=G\cap L_o$ for all $o$.

Knowing $o$,
the generators of $L_o\subseteq C$ are obviously decidable.
Hence, $L_o$ is effectively benign in \SAP $C$
according to Lemma~\ref{l:Benign}b).
Since the same applies to $G$, too,
Lemma~\ref{l:Benign}d) yields also
$W_o$ to be effectively benign in $C$;
and thus in $G$ as well by virtue of
Lemma~\ref{l:Benign}a).
\qed\end{proof}
%%%%%%%%%%%%%%%%%%%%%%%%%%%%%%%%%%%%%%%%%%%%%%%
\subsection{Putting It All Together} \label{s:Final}
So far, the index $n$ of the computational path $\gamma_n$
had been fixed. It will now run over $\IN$, so that
\begin{itemize}
\item[\textbullet]
  $n\mapsto\IA_n\subseteq\IR^{d(n)}$  
  denotes an enumerable and uniformly decidable decomposition of
  $\IH=\bigcup_{n\in\IN} \IA_n$;%
\item[\textbullet]
  $U_n:=\langle\bar w_{n,\bar r}:\bar r\in\IA_n\rangle\;\subseteq\; G$, $n\in\IN$,
  \quad where
\item[\textbullet]
  $G=\langle y;(x_{(i,s)})_{_{s\in\IR,i\in\IN}}\rangle$ 
  denotes a free \SAP group;
  \quad and
\item[\textbullet]
  $\bar w_{(n,r_1,\ldots,r_d)}=
  x^{-1}_{(d,r_d)}\cdots x^{-1}_{(1,r_1)}\cdot x^{-1}_{(0,n)}\cdot y
  \cdot x^{}_{(0,n)}\cdot x^{}_{(1,r_1)}\cdots x^{}_{(d,r_d)}$.
\end{itemize}
Observe how the index $n$ of the path $\gamma_n$ accepting $\IA_n$
is now encoded into the words generating $U_n$.
Theorem~\ref{t:Single} obviously carries over to this
minor modification, hence 
\begin{itemize}
\item[\textbullet]
  $U_n$ is effectively benign in $G$.
\end{itemize}
For given $n$, $\IA_n$ is decidable: simply evaluate
$\gamma_n$ on a given $\bar r$. This amounts
to \emph{uniform} decidability (Remark~\ref{r:Uniform}).
A brief review of Section~\ref{s:Single} reveals
all constructions to hold uniformly in $n$ so that in fact
\begin{itemize}
\item[\textbullet]
  $U_n$ is \emph{uniformly} effectively benign in $G$ in the sense of Definition~\ref{d:Benign}c).
\end{itemize}
It now follows from Lemma~\ref{l:Benign}f)
that $\langle\bigcup_n U_n\rangle\subseteq G$ is
effectively benign in $G$, too;
and so is 
\[ U\quad:=\quad
  \Big\langle\big\langle\bigcup\nolimits_n U_n\big\rangle\;\cup\; \big\langle (x_{(0,n)})_{_{n\in\IN}}\big\rangle\Big\rangle
   \;\cap\;\big\langle y ; (x_{(i,s)})_{_{s\in\IR, i\geq 1}}\big\rangle \]
by Lemma~\ref{l:Benign}b+d+e).
According to Remark~\ref{r:Benign}, membership to $U$ 
can thus be reduced to the word problem of some \SAP
group $K$. But, similar to the arguments in Lemmas~\ref{l:Single}
and \ref{l:Single2}, $U$ arises from $\bigcup_n U_n$ by eliminating
$x_{(0,n)}$ and replacing it with an existential quantifier
over $n$. Hence, $U$ equals
$\big\langle\{ \bar w_{\bar r}:\exists n : \bar r\in\IA_n\}\big\rangle$
by virtue of Fact~\ref{f:Nielsen}.
This concludes the proof of Theorem~\ref{t:Hardness}.
\qed

\medskip\noindent
More precisely, regarding Observation~\ref{o:Dimension}
and Footnote~\ref{f:Constants}, one arrives at the following
\begin{scholiumf}\footnotetext{A \aname{scholium} is ``\emph{a 
note amplifying a proof or course of reasoning,
as in mathematics}'' \cite{Dictionary}.}
To every \BSS machine $\IM$ semi-deciding some language
$\IP\subseteq\ri$, there exists an \SAP real group
$G=\langle X|R\rangle$ 
(in fact with $X\subseteq\IR\times\IN$)
to whose word problem the
membership in $\IP$ is reducible to.

The computation of this reduction requires no 
real constants. Moreover, deciding $X$ and $R$
is possible uniformly in (that is, given)
$\IM$. In particular, the description of $G$
requires no real constants other than those
present already in $\IM$.
\qed\end{scholiumf}
Since a Universal \BSS Machine does not need constants,
it follows
\begin{corollary}
The real Halting Problem $\IH$ is reducible to the
word problem of an \SAP group over $\IQ$!
\end{corollary}

%%%%%%%%%%%%%%%%%%%%%%%%%%%%%%%%%%%%%%%%%%%%%%%%%%%%%%%%%%%%%%%%%%%%%%%%
\section{Conclusions and Perspectives} \label{s:Conclusion}
In this paper we have introduced the class of \SAP real groups 
given as a quotient group of a free group and a normal subgroup. 
The free group was defined through a possibly uncountable set 
of generators \BSS-decidable in some fixed dimensional
space; the relations are similarly generated by a \BSS-decidable set.
We then considered the word problem for such groups: Given a finite
sequence of generators, decide whether this word is equivalent 
(with respect to the relations) to the unit element?

As main result of the paper it has been established that, on the one hand,
the word problem for an \SAP group is always
semi-decidable; while, on the other hand, there are \SAP
groups for which the above word problem is not only undecidable
but exactly as hard as the real Halting Problem. 

We believe our results to be an interesting step into the direction
of extending the \BSS theory into different areas of mathematics.
Many of the known computability and complexity results in the \BSS model 
are closely related to computational 
problems of semi-algebraic sets. Though these 
play an important role in our approach as well, the resulting problem
are located in the heart of computational group theory; their connection to 
semi-algebraic geometry is visible in the background only.  

\medskip
There are clearly a bunch of interesting questions to be investigated. 
We conclude by mentioning a few of them. They might hopefully serve as starting
point for a fruitful further research related to the topics studied in this paper.

Our construction yields a \BCSS-complete group with both
generators $X$ and relations $R$ being \BCSS-decidable.
\begin{myquestion}
Can we require the set of generators to be \emph{semi-algebraic}
rather than decidable?%
\end{myquestion}
Over complex numbers, every decidable set in some $\IC^N$ is also
algebraic \cite{Rossello}; however our proof makes heavy use of
$\IZ$ as a discrete component of $X$ and does not comply with
complex decidability.
\begin{myquestion}
How about a group with word problem \BCSS-complete
over $\IC$?
\end{myquestion}
In our approach, the relations $R$ seem crucial to live in $\ri$;
for instance in view of $U_{\gamma_n}$ (Definition~\ref{d:Single}) which includes
words $\bar w_{n;\bar r}$ of length $1+2d(n)$ unbounded in $n$.
\begin{myquestion} \label{q:InfiniteDim}
Can one restrict (not only the set of generators but also) the set of relations
to some finite-dimensional $\IR^M$? 
\end{myquestion}
To this end, it might be worth while exploiting that a \BSS machine
references data in fact not globally but through copy registers which change 
by at most one in each step; cf. Definition~\ref{defBCSS}.

It would furthermore be nice to have a real counterpart to the famous
Higman Embedding Theorem (Fact~\ref{f:Higman}):
\begin{myquestion}
Does every recursively presented real group admit a
(\BCSS-computable) embedding into an effectively presented one?
\end{myquestion}

\medskip\noindent
Special classes of discrete groups with \emph{decidable} word problem 
have been investigated 
with respect to the computational \emph{complexity} of this decision
\cite{Muller,Rees}. This looks promising to carry over to the reals;
for instance in form of
\begin{myquestion}
Can we find a class of groups whose word problem is (decidable and)
complete for a certain complexity class like
$\NP_{\IR}$ ?
\end{myquestion}
This would be interesting in order to extend the yet sparse list of 
known $\NP_{\R}$--complete problems. 

Finally, an entire bunch of interesting question results from inspecting 
further classical undecidability results in the new framework. 
We close here by just referring to the survey paper by Miller \cite{Miller} 
in which a lot of related issues are discussed.

%%%%%%%%%%%%%%%%%%%%%%%%%%%%%%%%%%%%%%%%%%%%%%%%%%%%%%%%%%%%%%%%%%%%%%%%

\end{document}